\documentclass[pre,aps,groupaddress,showpacs,onecolumn]{revtex4-1}
\usepackage{epsfig,amsmath,graphicx,amssymb,overpic}
\usepackage{graphicx}
\usepackage{dcolumn}
\usepackage{bm}
\usepackage{graphicx}
\usepackage{subfigure,color}

\def\be{\begin{equation}}
\def\ee{\end{equation}}
\def\bee{\begin{eqnarray}}
\def\ene{\end{eqnarray}}
\def\bes{\begin{subequations}}
\def\ees{\end{subequations}}

\begin{document}

\title{Higher-order vector discrete rogue-wave states in the coupled
Ablowitz-Ladik equations: exact solutions and stability}
\author{Xiao-Yong Wen$^{1,2}$}
\author{Zhenya Yan$^1$}
\email{zyyan@mmrc.iss.ac.cn}
\author{Boris A. Malomed$^{3}$}
\affiliation{\vspace{0.1in}
    $^1$Key Laboratory of Mathematics Mechanization, Institute of Systems Science, AMSS, Chinese Academy of Sciences, Beijing 100190, China\\
 $^2$Department of Mathematics, School of Applied Science, Beijing Information
    Science and Technology University, Beijing  100192, China \\
 $^3$Department of Physical Electronics, School of Electrical Engineering,  Tel Aviv University, Tel Aviv 59978, Israel}

\vspace{0.2in}

\begin{abstract}

 {\it This paper has been accepted for publication in Chaos (2016).}

\vspace{0.2in} \baselineskip=15pt

An integrable system of two-component nonlinear Ablowitz-Ladik (AL)
equations is used to construct complex rogue-wave (RW) solutions in an
explicit form. First, the modulational instability of continuous waves is
studied in the system. Then, new higher-order discrete two-component RW
solutions of the system are found by means of a newly derived discrete
version of a generalized Darboux transformation. Finally, perturbed
evolution of these RW states is explored in terms of systematic simulations,
which demonstrates that tightly and loosely bound RWs are, respectively,
nearly stable and strongly unstable solutions.

\vspace{0.3in}

\textbf{Since the rogue waves (RWs) were identified from observations in the
deep ocean, they have been shown to appear in many fields of nonlinear
science, such as nonlinear optics, Bose-Einstein condensates, plasma
physics, and even financial dynamics. The RWs can be produced as complex
solutions of various nonlinear wave equations -- first of all, the nonlinear
Schr\"{o}dinger (NLS) equation with the self-focusing cubic nonlinearity. In
the framework of the integrable NLS equation, RW solutions can be generated
from trivial ones by means of the Darboux transform (DT). Generally, a
majority of theoretical studies of the RWs addressed models based on
continuous partial differential equations, except for a few results which
produced discrete RWs in integrable discrete systems, including the
Ablowitz-Ladik (AL) equation (an integrable discretization of the NLS
equation) and discrete Hirota equation. A natural possibility is to extend
the search for RW states to coupled (first of all, two-component) systems of
nonlinear wave equations. The present work aims to construct complex
(higher-order) discrete RW solutions in an integrable system of two AL
equations with self-attractive nonlinear terms. For this purpose, we derive
a newly generalized DT method for the system, and then use it to generate
higher-order discrete RW states. Then, perturbed evolution of these RWs is
investigated, in a systematic form, by means of numerical simulations. It is
concluded that the RW states with a tightly bound structure are
quasi-stable, while their loosely bound counterparts are subject to a strong
instability.}
\end{abstract}

\maketitle

\affiliation{\vspace{0.1in}
    $^1$Key Laboratory of Mathematics Mechanization, Institute of Systems Science, AMSS, Chinese Academy of Sciences, Beijing 100190, China\\
 $^2$Department of Mathematics, School of Applied Science, Beijing Information
    Science and Technology University, Beijing  100192, China \\
 $^3$Department of Physical Electronics, School of Electrical Engineering,  Tel Aviv University, Tel Aviv 59978, Israel}

%\email{wenxiaoyong@amss.ac.cn}

\baselineskip=15pt

\section{Introduction}

Cubic nonlinear Schr\"{o}dinger (NLS) equations model a great variety of
phenomena in many fields of nonlinear science, including nonlinear optics,
water waves, plasma physics, superconductivity, Bose-Einstein
condensates~(BECs), and even financial dynamics~\cite{soliton}-\cite{f3}. A
class of unstable but physically meaningful solutions of the cubic NLS
equation represents rogue waves (RWs), which spontaneously emerge due to the
modulational instability (MI) of continuous-wave (CW) states and then
disappear \cite{RW1,RW2}. Recently, some new RW structures generated by the
self-focusing NLS equation and some of its extensions were found, see, e.g., Refs.~\cite{f3}-%
\cite{yuan16}. Most of these works studied continuous models, except
for a few works which addressed the single-component Ablowitz-Ladik (AL)
equation \cite{Ablowitz} and discrete Hirota equation, using the modified
bilinear method~\cite{nail1d, nail2d,yang14} and the generalized discrete
Darboux transformation~\cite{wen-yan}. Being a fundamentally
important mathematical model, the AL equation does not find many physical
realizations, but it can be implemented, in principle, as a model of arrayed
optical waveguides \cite{optics}, and as a mean-field limit of generalized
Bose-Hubbard model for BECs trapped in a deep optical-lattice potential \cite%
{Kuba}.

There also exist interesting coupled discrete nonlinear wave systems -- in
particular, coupled AL equations:
\begin{equation}
\begin{array}{ll}
R_{n,t}=-i (\sigma+ |R_{n}|^2)({S_{n}}^{\ast }+{S_{n-1}}^{\ast }),\vspace{%
0.1in} &  \\
S_{n,t}=i (\sigma +|S_{n}|^2)({R_{n+1}}^{\ast }+{R_{n}}^{\ast }), &
\end{array}
\label{nls}
\end{equation}%
where $n$ denotes the discrete spatial variable, and $t$ stands for time, $%
R_n\equiv R_{n}(t)$ and $S_n\equiv S_{n}(t)$ are the lattice dynamical
variables, $\left\{ R_{n,t},S_{n,t}\right\} \equiv d\left\{
R_{n},S_{n}\right\} /dt$, and the asterisk stands for the complex conjugate,
while $\sigma =+1$ and $-1$ correspond, respectively, to the self-focusing
and defocusing signs of the nonlinearity. System (\ref{nls}) is an
integrable system, which represents a special two-field reduction of the
four-field AL equation introduced and solved by means of the
inverse-scattering transform in Ref.~\cite{Ablowitz} (a more general
nonintegrable form of coupled AL equations was introduced in Ref. \cite{MY}%
). $N$-soliton solutions of Eq. (\ref{nls}), obtained in terms of
determinants, and the derivation of infinitely many conservation laws for
this system in the defocusing case, with $\sigma =-1$, by means of the $N$%
-fold Darboux transform (DT),~were reported in Ref. \cite{wen1}. More
recently, discrete vector (two-component) RW solutions of the coupled AL
equations,
\begin{equation}
i\psi _{n,t}^{(j)}+\frac{1}{2h^{2}}(\psi _{n-1}^{(j)}+\psi
_{n+1}^{(j)}-2\psi _{n}^{(j)})+\frac{1}{2}(\psi _{n-1}^{(j)}+\psi
_{n+1}^{(j)})(|\psi _{n}^{(j)}|^{2}+|\psi _{n}^{(3-j)}|^{2})\quad (j=1,2),
\label{dnls}
\end{equation}%
were produced too, using a special transformation~\cite{nail3}.

To the best of our knowledge, higher-order discrete RW solutions of system (%
\ref{nls}) have not been investigated before. A new generalized $(n,N-n)$%
-fold DT technique, generating higher-order RWs, was developed for some
\textit{continuous} nonlinear wave equations~\cite{yan-wen-pre,yan-wen-chaos}%
, but it is still an issue how to apply a similar technique to \textit{%
discrete} nonlinear equations. The present paper addresses this issue for
discrete coupled equations (\ref{nls}). The key feature of such a technique
is the fact that the Lax pair associated with the integrable nonlinear
differential-difference equations is covariant under the gauge
transformation~\cite{wen1,wen2}.

The rest of the paper is arranged as follows. In Sec. II, we investigate the
MI in the framework of Eq.~(\ref{nls}), starting with its CW solution. In
Sec. III, based on its Lax pair, a new $N$-fold DT for Eq.~(\ref{nls}) is
constructed, which can be used to derive $N$-soliton solutions from the zero
seed solution, and $N$-breather solutions from the CW seed solutions of Eq.~(%
\ref{nls}). Then we present a novel idea, allowing one to derive discrete
versions of the generalized $(1,N-1)$-fold (using one spectral parameter)
and $(M,N-M)$-fold (using $M<N$ spectral parameters) DTs for Eq.~(\ref{nls}%
), by means of the Taylor expansion and a limit procedure related to the $N$%
-fold Darboux matrix. In Sec. IV, the generalized $(1,N-1)$-fold D is used
to find discrete higher-order vector RW solutions of Eq.~(\ref{nls}). We
also analyze RW structures and their dynamical behavior by means of
numerical simulations. The paper is concluded by Sec. V.

\section{Modulational instability of continuous waves}

The simplest solution of system (\ref{nls}) with $\sigma =1$ (the
self-focusing nonlinearity) is the CW state:
\begin{equation}
R_{0}(n,t)=ice^{2i(c^2+1)t},\quad S_{0}(n,t)=ice^{-2i(c^{2}+1)t}  \label{s1}
\end{equation}%
with the real amplitude $c\not=0$. Investigation of the MI of this state is
the first step towards the construction of the RW modes, as they are
actually produced by the instability of the CW background \cite{RW1,RW2}.

We begin the analysis of the MI, taking the perturbed form of the CW as
\begin{equation}
R_{n}(t)=[ic+\varepsilon r_{n}(t)]e^{2i(c^{2}+1)t},\ \ \
S_{n}(t)=[ic+\varepsilon s_{n}^{\ast}(t)]e^{-2i(c^{2}+1)t},  \label{sp}
\end{equation}%
where $\varepsilon $ is an infinitesimal amplitude of the perturbation. The
substitution of wave form (\ref{sp}) into system (\ref{nls}) yields a
linearized system,
\begin{equation}
\begin{array}{ll}
r_{n,t}+2i\left(r_{n}+c^2r_n^{*}\right)+
i\left(c^{2}+1\right)\left(s_{n}+s_{n-1}\right) =0,\vspace{0.1in} &  \\
s_{n,t}+2i\left(s_{n}+c^2s_n^{*}\right)+i\left(
c^{2}+1\right)\left(r_{n+1}+r_{n}\right) =0. &
\end{array}
\label{le}
\end{equation}

\begin{figure}[tbp]
\begin{center}
{\includegraphics[scale=0.3]{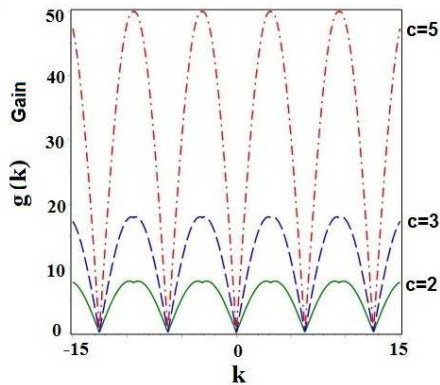}}
\end{center}
\par
\vspace{-0.25in}
\caption{{\protect\small (color online). Gain spectra of the modulation
instability for different CW amplitudes are displayed by means of positive
growth rate $g(k)$, as given by Eq. (\protect\ref{g}) with sign $+$.
Different lines correspond to values of the CW amplitude $c=2$ (solid), $c=3$
(dashed), and $c=5$ (dotted-dashed), as indicated in the figure.}}
\label{gain-mi}
\end{figure}

To analyze the MI on the basis of the complex linearized equation (\ref{le}), each component of the perturbation may be split into real and imaginary
parts: $r_{n}(t)\equiv r_{1n}(t)+ir_{2n}(t),~s_{n}(t)\equiv s_{1n}(t)+is_{2n}(t)$,~which transform the system of two complex equations (\ref{le}) into a system
of four real equations:
\begin{equation}
\begin{array}{l}
r_{1n,t}+2\left( c^{2}-1\right)
r_{2n}-(c^{2}+1)\left(s_{2n}+s_{2,n-1}\right) =0, \vspace{0.1in} \\
r_{2n,t}+2\left( c^{2}+1\right)
r_{1n}+(c^{2}+1)\left(s_{1n}+s_{1,n-1}\right) =0, \vspace{0.1in} \\
s_{1n,t}+2\left( c^{2}-1\right)
s_{2n}-(c^{2}+1)\left(r_{2,n+1}+r_{2n}\right) =0, \vspace{0.1in} \\
s_{2n,t}+2\left( c^{2}+1\right)
s_{1n}+(c^{2}+1)\left(r_{1,n+1}+r_{1n}\right) =0.%
\end{array}
\label{real}
\end{equation}%
Solutions to these real equations may be sought for in a formally complex
form,
\begin{equation}
\left\{ r_{1n}(t),\,r_{2n}(t),\,s_{1n}(t),\,s_{2n}(t)\right\} =\left\{
r_{1n}^{(0)},\,r_{2n}^{(0)},\,s_{1n}^{(0)},\,s_{2n}^{(0)}\right\} e^{gt+ikn},
\label{real2}
\end{equation}%
where $g$ is the MI gain (generally, this eigenvalue may be complex), and $k$
an arbitrary real wavenumber of the small perturbation, while $%
r_{1n}^{(0)},\,r_{2n}^{(0)},\,s_{1n}^{(0)},\,s_{2n}^{(0)}$ are constant
amplitudes of the perturbation eigenmode. The substitution of this ansatz in
Eq.~(\ref{real2}) into system (\ref{real}) produces an MI dispersion
equation in the form of the determinant,
\begin{equation}
\left\vert
\begin{array}{cccc}
g & 2(c^{2}-1) & 0 & -(c^{2}+1)(1+e^{-ik})\vspace{0.1in} \\
2(c^{2}+1) & g & (c^{2}+1)(1+e^{-ik}) & 0\vspace{0.1in} \\
0 & -(c^{2}+1)(1+e^{ik}) & g & 2(c^{2}-1)\vspace{0.1in} \\
(c^{2}+1)(1+e^{ik}) & 0 & 2(c^{2}+1) & g%
\end{array}%
\right\vert =0,  \label{det}
\end{equation}%
which leads to an explicit dispersion relation:
\begin{equation}
g^{2}=2(c^{2}+1)\left[ c^{2}-(c^{2}+1)\cos k-3\pm 2\sqrt{2(1+\cos k)}\right]
.  \label{g}
\end{equation}%
Therefore, the MI occurs when expression (\ref{g}) is positive, the branch
with sign $+$ being always dominant. In particular, it is easy to check that
the MI condition $g^{2}>0$ holds for \emph{all} $k$, provided that the CW
amplitude is large enough, \textit{viz}., $c^{2}>1$. To illustrate this
property, Fig.~\ref{gain-mi} displays the positive root $g(k)$ for $c=2,3,$
and $5$. Furthermore, at point
\begin{equation}
\cos k=2\left(c^{2}+1\right) ^{-2}-1  \label{cos}
\end{equation}
(note that it obeys constraint $\left\vert \cos k\right\vert <1$), Eq. (\ref%
{g}) yields $g_{\max }^{2}=4c^{4}$, hence the MI takes place at \emph{all}
finite values of amplitude $c$. In fact, $g_{\max }^{2}$ is the largest
value attained by $g^{2}\left( \cos k\right) $ at any $c$, i.e., Eq. (\ref%
{cos}) determines wavenumbers of the dominant MI perturbation eigenmodes, $%
k=\pm \arccos \left[ 2\left(c^{2}+1\right) ^{-2}-1\right] +2m\pi$ with
integer $m$.

\section{Generalized Darboux transform}

\subsection{The Lax pair and $N$-fold Darboux transform}

The Lax pair of system (\ref{nls}) is written in the known form~\cite{wen1}:
\begin{equation}
E\varphi _{n}=U_{n}\varphi _{n}, \qquad U_n= \left(%
\begin{array}{cc}
\lambda +R_{n}S_{n} & -\sigma {R_{n}}^{\ast }+\lambda^{-1}S_{n}\vspace{0.1in}
\\
R_{n}-\sigma \lambda {S_{n}}^{\ast } & \lambda^{-1}+(R_{n}S_{n})^{\ast }%
\end{array}%
\right),  \label{lax1}
\end{equation}
\begin{equation}
\varphi _{n,t}=V_{n}\varphi _{n}, \qquad V_n=i\left(%
\begin{array}{cc}
-\lambda +(R_{n}S_{n-1})^{\ast } & \quad \sigma {R_{n}}^{\ast}+%
\lambda^{-1}S_{n-1}\vspace{0.1in} \\
R_{n}+\sigma \lambda {S_{n-1}}^{\ast } & \lambda^{-1}-R_{n}S_{n-1}%
\end{array}%
\right),\ \   \label{lax2}
\end{equation}
where $\lambda $ is a constant iso-spectral parameter, $E$ is the shift
operator defined by $Ef_{n}=f_{n+1},\ E^{-1}f_{n}=f_{n-1}$, and $%
\varphi_{n}=\varphi_{n}(t)=(\phi _{n}(t),\psi _{n}(t))^{T}$ (with $T$
denoting the transpose of the vector or matrix) is an eigenfunction vector.
It is easy to verify that the compatibility condition, $%
U_{n,t}+U_nV_n-V_{n+1}U_n=0$, of Eqs.~(\ref{lax1}) and (\ref{lax2}) is
precisely tantamount to system (\ref{nls}).

In what follows, we construct a new $N$-fold DT for system (\ref{nls}),
which is different from one reported in Ref.~\cite{wen1}. For this purpose,
we introduce the following gauge transformation:
\begin{equation}
\widetilde{\varphi }_{n}=T_n\varphi _{n}\ ,  \label{gauge}
\end{equation}%
where $T_n$ is an $2\times 2$ Darboux matrix to be determined, $\widetilde{%
\varphi }_{n}$ is required to satisfy Eqs. (\ref{lax1}) and (\ref{lax2})
with $U_{n}$ and $V_{n}$ replaced, respectively, by $\widetilde{U_{n}}$ and $%
\widetilde{V_{n}}$, i.e.,
\begin{equation}
E\widetilde{\varphi }_{n}=\widetilde{U}_{n}\widetilde{\varphi }_{n},\quad
\widetilde{\varphi }_{n,t}=\widetilde{V}_{n}\widetilde{\varphi }_{n},
\label{laxg}
\end{equation}%
with $\widetilde{U}_{n}=U_{n}|_{R_{n}\rightarrow \widetilde{R}%
_{n},\,S_{n}\rightarrow \widetilde{S}_{n}}$ and $\widetilde{V}%
_{n}=V_{n}|_{R_{n}\rightarrow \widetilde{R}_{n},\,S_{n}\rightarrow
\widetilde{S}_{n}}$. It follows from Eq.~(\ref{gauge}) and (\ref{laxg}) that
one has constraints
\begin{equation}
\widetilde{U}_{n}=T_{n+1}U_{n}{T_n}^{-1},\qquad \widetilde{V}%
_{n}=(T_{n,t}+T_{n}V_{n}){T_n}^{-1},  \label{dm-c}
\end{equation}%
which lead to the DT of system (\ref{nls}).

The Darboux matrix $T_{n}$, which determines the gauge transformation (\ref%
{gauge}) is a key element in constructing the DT of Eq. (\ref{nls}). We
consider the matrix in the form of
\begin{equation}
T_n(\lambda )=\left(
\begin{array}{cc}
\lambda ^{N}+\sum\limits_{j=0}^{N-1}a_{n}^{(j)}\lambda ^{j} &
\sum\limits_{j=0}^{N-1}b_{n}^{(j)}\lambda ^{j}\vspace{0.1in} \\
-\sigma \sum\limits_{j=0}^{N-1}{b_{n}^{(j)}}^{\ast }\lambda ^{N-j} & \quad
1+\sum\limits_{j=0}^{N-1}{a_{n}^{(j)}}^{\ast }\lambda ^{N-j}%
\end{array}\right),
\label{Tmatrix}
\end{equation}
where $N$ is a non-negative integer, and $a_{n}^{(j)},b_{n}^{(j)}%
\,(j=0,1,2,...,N-1)$ are functions of $n$ and $t$ determined by the
following linear algebraic system:
\begin{equation}
T_n(\lambda )\varphi (\lambda )|_{\lambda =\lambda _{k}}=0\quad (k=1,2,...,N)
\label{ab2}
\end{equation}%
where $\varphi (\lambda )=(\phi (\lambda ),\psi (\lambda ))^{T}$ is a basic
solution of Eqs.~(\ref{lax1}) and (\ref{lax2}). $N$ parameters $\lambda
_{k}\ (\lambda _{k}\neq \lambda _{j}$ for $k\neq j)$ must be chosen so that
the determinant of the coefficients for system~(\ref{ab2}) is different from
zero, hence, $a_{n}^{(j)}$ and $b_{n}^{(j)}$ are uniquely determined by Eq.~(%
\ref{ab2}). It follows from Eqs.~(\ref{dm-c}) and (\ref{ab2}) that we have
the following theorem: \newline

\noindent \textbf{Theorem 1.} \thinspace\ \textit{Let $\varphi _{j}(\lambda
_{j})=(\phi _{j},\psi _{j})^{T}$ be $N$ distinct column-vector solutions of
the problem for spectral parameters $\lambda _{j}\,(j=1,2,..,N)$, based on
Eqs. (\ref{lax1}) and (\ref{lax2}), and $R_{0}(n,t),\, S_{0}(n,t)$ is a seed
solution of system (\ref{nls}); then, the new $N$-fold DT for system (\ref%
{nls}) is given by
\begin{equation}
\hspace{2cm}\widetilde{R}_{N}(n,t)=\dfrac{R_{0}(n,t)+\sigma {b_{n}^{(N-1)}}%
^{\ast }}{a_{n}^{(0)}},\ \ \ \widetilde{S}_{N}(n, t)=\sigma {%
b_{n+1}^{(0)}+S_{0}(n,t)a_{n+1}^{(0)}},  \label{dt}
\end{equation}%
with $a_{n}^{(0)}=\dfrac{\Delta a_{n}^{(0)}}{\Delta _{N}},b_{n}^{(0)}=\dfrac{%
\Delta b_{n}^{(0)}}{\Delta _{N}},b_{n}^{(N-1)}=\dfrac{\Delta b_{n}^{(N-1)}}{%
\Delta _{N}},$
\begin{equation}
\Delta _{N}=\left\vert
\begin{array}{cccccccc}
\lambda _{1}^{N-1}\phi _{1} & \lambda_{1}^{N-2}\phi _{1} & \ldots \quad \quad
& \phi _{1} & {\lambda _{1}}^{N-1}\psi _{1} & \lambda _{1}^{N-2}\psi_{1} &
\ldots & \psi _{1}\vspace{0.1in} \\
\lambda _{2}^{N-1}\phi _{2} & \lambda _{2}^{N-2}\phi _{2} & \ldots \quad\quad
& \phi _{2} & \lambda _{2}^{N-1}\psi _{2} & \lambda _{2}^{N-2}\psi_{2} &
\ldots & \psi _{2}\vspace{0.1in} \\
\ldots & \ldots & \ldots \quad \quad & \ldots & \ldots & \ldots & \ldots &
\ldots \vspace{0.1in} \\
\lambda _{N}^{N-1}\phi _{N} & \lambda _{N}^{N-2}\phi _{N} & \ldots \quad
\quad & \phi _{N} & \lambda _{N}^{N-1}\psi _{N} & \lambda _{N}^{N-2}\psi _{N}
& \ldots & \psi _{N}\vspace{0.1in} \\
\lambda _{1}^*\psi _{1}^* & \lambda _{1}^{\ast 2}\psi _{1}^* & \ldots \quad \quad
& \lambda _{1}^{\ast N}\psi _{1}^* & -\sigma \lambda _{1}^*\phi _{1}^* & -\sigma
\lambda _{1}^{\ast 2}\phi _{1}^* & \ldots & -\sigma \lambda _{1}^{\ast N}\phi_{1}^*\vspace{0.1in} \\
\lambda _{2}^*\psi _{2}^* & \lambda _{2}^{\ast 2}\psi _{2}^* & \ldots \quad \quad
& \lambda _{2}^{\ast N}\psi _{2}^* & -\sigma \lambda _{2}^*\phi _{2}^* & -\sigma
\lambda _{2}^{\ast 2}\phi _{2}^* & \ldots & -\sigma \lambda _{2}^{\ast N}\phi_{2}^*\vspace{0.1in} \\
\ldots & \ldots & \ldots \quad \quad & \ldots & \ldots & \ldots & \ldots &
\ldots \vspace{0.1in} \\
\lambda _{N}^*\psi _{N}^* & \lambda _{N}^{\ast 2}\psi _{N}^* & \ldots \quad \quad
& \lambda _{N}^{\ast N}\psi _{N}^* & -\sigma \lambda _{N}^*\phi _{N}^* & -\sigma
\lambda _{N}^{\ast 2}\phi _{N}^* & \ldots & -\sigma \lambda _{N}^{\ast N}\phi_{N}^*
\end{array}
\right\vert ,  \notag
\end{equation}%
where $\Delta a_{n}^{(0)}$, $\Delta b_{n}^{(0)}$ and $\Delta b_{n}^{(N-1)}$
are given by determinant $\Delta _{N}$, replacing its $N$-th, $(2N)$-th and $%
(N+1)$-th columns by the column vector $(-\lambda _{1}^{N}\phi _{1}$, $%
-\lambda _{2}^{N}\phi _{2}$,$\cdots $, $-\lambda _{N}^{N}\phi _{N}$, $-\psi
_{1}^{{*}}$, $-\psi _{2}^{{*}}$,$\cdots $, $-\psi _{N}^{{*}})^{\mathrm{T}}$,
while $\Delta a_{n+1}^{(0)}$ and $\Delta b_{n+1}^{(0)}$ are obtained from $%
\Delta a_{n}^{(0)}$ and $\Delta b_{n}^{(0)}$, replacing $n$ by $n+1$.}

The proof of this theorem can be performed following the lines of Refs.~\cite%
{wen1,wen2}, and this $N$-fold DT includes the known DT~\cite{wen1}. The $N$%
-fold DT with the zero seed solutions, $R_{0}=S_{0}=0$ (or the CW, alias
plane wave, solutions), can be used to construct multi-soliton solutions (or
multi-breather solutions) of Eq.~(\ref{nls}). This is not the main objective
of the present work. Actually, our aim is to extend the $N$-fold DT into a
generalized $(M,N-M)$-fold DT such that multi-RW solutions can be found in
terms of the determinants defined above for solutions of Eq.~(\ref{nls}).

\subsection{Generalized $(1,N-1)$-fold Darboux transforms}

Here we consider the Darboux matrix (\ref{Tmatrix}) with a single spectral
parameter, $\lambda =\lambda _{1}$, rather than $N>1$ distinct parameters $%
\lambda _{k}\,$. Then, the respective equation (\ref{ab2}),
\begin{equation}
T_n(\lambda _{1})\varphi (\lambda _{1})=0,  \label{teq}
\end{equation}%
gives rise to two algebraic equations for $2N$ functions $a_{n}^{(j)}$ and $%
b_{n}^{(j)}\,(j=0,1,...,N-1)$. To determine them, we need to find additional
$2(N-1)$ equations for $a_{n}^{(j)}$ and $b_{n}^{(j)}$. To generate such
extra equations from Eq.~(\ref{teq}), we consider the Taylor expansion of $%
T_n(\lambda _{1})\varphi (\lambda _{1})\big|_{\{\lambda _{1}\rightarrow
\lambda _{1}+\varepsilon \}}=T_n(\lambda _{1}+\varepsilon )\varphi (\lambda
_{1}+\varepsilon )$ around $\varepsilon =0$ in the form of
\begin{equation}
T_n(\lambda _{1}+\varepsilon )\varphi (\lambda _{1}+\varepsilon
)=\sum_{k=0}^{N-1}\sum\limits_{j=0}^{k}T^{(j)}(\lambda _{1})\varphi
^{(k-j)}(\lambda _{1})\varepsilon ^{k}+\mathcal{O}(\varepsilon ^{N-1}),\quad
\label{teqg}
\end{equation}%
and demand that all coefficients in front of $\varepsilon^{s}\,(s=0,1,2,...,N-1)$ are equal to zero, which yields
\begin{equation}
\begin{array}{r}
T_n^{(0)}(\lambda _{1})\varphi ^{(0)}(\lambda _{1})=0,\vspace{0.1in} \\
T_n^{(0)}(\lambda _{1})\varphi ^{(1)}(\lambda _{1})+T_n^{(1)}(\lambda
_{1})\varphi ^{(0)}(\lambda _{1})=0,\vspace{0.1in} \\
T_n^{(0)}(\lambda _{1})\varphi ^{(2)}(\lambda _{1})+T_n^{(1)}(\lambda
_{1})\varphi ^{(1)}(\lambda _{1})+T_n^{(2)}(\lambda _{1})\varphi
^{(0)}(\lambda _{1})=0,\vspace{0.1in} \\
\qquad \cdots \cdots ,\qquad \qquad \vspace{0.1in} \\
\sum\limits_{j=0}^{N-1}T_n^{(j)}(\lambda _{1})\varphi ^{(N-1-j)}(\lambda
_{1})=0.%
\end{array}
\label{nls1sys}
\end{equation}%
where
\begin{eqnarray}  \label{tderivative}
T_n^{(k)}(\lambda _{1})=\frac{1}{k!}\frac{\partial ^{k}}{\partial \lambda
_{1}^{k}}T(\lambda _{1}),\quad \varphi ^{(k)}(\lambda _{1})=\frac{1}{k!}%
\frac{\partial ^{k}}{\partial \lambda _{1}^{k}}\varphi (\lambda _{1})=\left(
\frac{1}{k!}\frac{\partial ^{k}}{\partial \lambda _{1}^{k}}\phi (\lambda
_{1}),\,\frac{1}{k!}\frac{\partial ^{k}}{\partial \lambda _{1}^{k}}\psi
(\lambda _{1})\right) ^{\mathrm{T}}\quad (k=0,1,2,...).
\end{eqnarray}

The first sub-system in system (\ref{nls1sys}) is just Eq.~(\ref{teq}), as
required. Therefore, system (\ref{nls1sys}) contains $2N$ algebraic
equations for $2N$ unknowns $a_{n}^{(j)}$ and $b_{n}^{(j)}\,(j=0,1,...,N-1)$%
. When eigenvalue $\lambda _{1}$ is suitably chosen, so that the determinant
of coefficients of system~(\ref{nls1sys}) does not vanish, the Darboux
matrix $T_n$ given by Eq.~(\ref{Tmatrix}) is uniquely determined by the new
system~(\ref{nls1sys}) (cf. the previous condition given by Eq.~(\ref{ab2})).

The proof of Theorem 1 only uses Eqs.~(\ref{gauge}), (\ref{dm-c}), and (\ref%
{Tmatrix}), but not Eq.~(\ref{ab2})~\cite{wen2}, therefore, if we replace
system (\ref{ab2}) by (\ref{nls1sys}), which are used to determine $%
a_{n}^{(j)}$ and $b_{n}^{(j)}$ in Darboux matrix $T_n$, then Theorem 1 also
holds for Darboux matrix (\ref{Tmatrix}) with $a_{n}^{(j)}$ and $b_{n}^{(j)}$
determined by system (\ref{nls1sys}), cf. Ref.~\cite{wen2}. Thus, when $%
a_{n}^{(j)}$ and $b_{n}^{(j)}$ in the Darboux matrix $T_n$ are chosen as the
new functions, we can derive new solutions by using DT (\ref{dt}) with the
single eigenvalue $\lambda =\lambda _{1}$. We call Eqs.~(\ref{dt}) and (\ref%
{gauge}), associated with new functions $a_{n}^{(j)}$ and $b_{n}^{(j)}$
determined by system (\ref{nls1sys}), as a generalized $(1,N-1)$-fold DT,
which leads to the following theorem for system (\ref{nls}).

\vspace{0.2in}

\noindent\textbf{Theorem 2.} \textit{Let $\varphi (\lambda _{1})=(\phi
(\lambda _{1}),\psi (\lambda _{1}))^{\mathrm{T}}$ be a column
vector-solution of the problem for spectral parameter $\lambda _{1}$, based
on Eqs. (\ref{lax1}) and (\ref{lax2}), and $R_{0}(n,t), S_{0}(n, t)$ is a
seed solution of system (\ref{nls}); then, the generalized $(1,N-1)$-fold DT
for Eq.~(\ref{nls}) is given by
\begin{equation}
\hspace{2cm}\widetilde{R}_{N}(n,t)=\dfrac{R_{0}(n,t)+\sigma {b_{n}^{(N-1)}}%
^{\ast }}{a_{n}^{(0)}},\ \ \ \widetilde{S}_{N}(n,t)=\sigma {%
b_{n+1}^{(0)}+S_{0}(n,t)a_{n+1}^{(0)}},
\end{equation}%
with $a_{n}^{(0)}=\dfrac{\Delta a_{n}^{(0)}}{\Delta _{N}^{\epsilon }},\,
b_{n}^{(0)}=\dfrac{\Delta b_{n}^{(0)}}{\Delta _{N}^{\epsilon }},\,
b_{n}^{(N-1)}=\dfrac{\Delta b_{n}^{(N-1)}}{\Delta _{N}^{\epsilon }}$, $%
\Delta _{N}^{\epsilon }=\left\vert (\Delta _{j,s})_{2N\times 2N}\right\vert,$
\begin{equation}
\Delta _{j,s}=%
\begin{cases}
\sum\limits_{k=0}^{j-1}C_{N-s}^{k}{\lambda _{1}}^{(N-s-k)}{\phi ^{(j-1-k)}}%
\quad \mathrm{for}\quad 1\leq j,\,s\leq N,\vspace{0.1in} \\
\sum\limits_{k=0}^{j-1}C_{2N-s}^{k}{\lambda _{1}}^{(2N-s-k)}{\psi ^{(j-1-k)}}%
\quad \mathrm{for}\quad 1\leq j\leq N,\,N+1\leq s\leq 2N,\vspace{0.1in} \\
\sum\limits_{k=0}^{j-(N+1)}C_{s}^{k}{\lambda _{1}}^{\ast (s-k)}{\psi
^{*(j-N-1-k)}}\quad \mathrm{for}\quad N+1\leq j\leq 2N,\,1\leq s\leq N,%
\vspace{0.1in} \\
-\sigma \sum\limits_{k=0}^{j-(N+1)}C_{s-N}^{k}{\lambda _{1}}^{\ast (s-N-k)}{%
\phi ^{*(j-N-1-k)}}\quad \mathrm{for}\quad N+1\leq j,\,s\leq 2N,%
\end{cases}
\notag
\end{equation}%
with $\Delta a_{n}^{(0)}$, $\Delta b_{n}^{(0)}$  and $\Delta b_{n}^{(N-1)}$ given by determinant $%
\Delta _{N}^{\epsilon }$, replacing its $N$-th, $(2N)$-th and $(N+1)$-th columns by the column
vector $b=(b_{j})_{2N\times 1}$ with
\begin{equation}
b_{j}=
\begin{cases}
-\sum\limits_{k=0}^{j-1}C_{N}^{k}{\lambda _{1}}^{(N-k)}{\phi ^{(j-1-k)}}%
\quad \mathrm{for}\quad 1\leq j\leq N\vspace{0.1in} \\
-{\psi ^{*(j-N-1)}}\quad \mathrm{for}\quad N+1\leq j\leq 2N%
\end{cases}  \notag
\end{equation}%
Here, $\Delta a_{n+1}^{(0)}$ and $\Delta b_{n+1}^{(0)}$ are obtained from $%
\Delta a_{n}^{(0)}$ and $\Delta b_{n}^{(0)}$, replacing $n$ with $n+1$.}

Notice that, in the symbol of the generalized $(1,N-1)$-fold DT, integer $1$
refers to the number of the spectral parameters, and $N-1$ indicates the
order of the highest derivative of the Darboux matrix $T$ in system (\ref%
{nls1sys}) (or the order of the highest derivative of the vector
eigenfunction $\varphi $ in system (\ref{nls1sys})).

\subsection{Generalized $(M,N-M)$-fold Darboux transforms}

The generalized $(1,N-1)$-fold DT, which we have constructed using the
single spectral parameter, $\lambda =\lambda _{1}$, and the highest-order
derivatives of $T_n(\lambda _{1})$ or $\varphi (\lambda _{1})$ with $%
m_{1}=N-1$, can be extended to include $M<N$ spectral parameters $\lambda
_{i}\,(i=1,2,...,M)$ and the corresponding highest-order derivative orders, $%
m_{i}\,\ (m_{i}=0,1,2,...)$, where non-negative integers $M,\,m_{i}$ are
required to satisfy constraint $N=M+\sum_{i=1}^{M}m_{i}$, where $N$ is the
same as in Darboux matrix $T_n$ (\ref{Tmatrix}). In the following, we
consider the Darboux matrix (\ref{Tmatrix}) and eigenfunctions $\varphi
_{i}(\lambda _{i})\,(i=1,2,...,M)$, which are solutions of the linear
spectral problem for parameters $\lambda _{i}$, based on Eqs.~(\ref{lax1})
and (\ref{lax2}), along with the seed solution $R_{0},S_{0}$ of Eq.~(\ref%
{nls}). Thus we have
\begin{equation}
T_n(\lambda _{i}+\varepsilon )\varphi _{i}(\lambda
_{i}+\varepsilon)\!\!=\!\!\sum_{k=0}^{+\infty
}\sum\limits_{j=0}^{k}T_n^{(j)}(\lambda_{i})\varphi _{i}^{(k-j)}(\lambda
_{i})\varepsilon ^{k},  \label{kpg}
\end{equation}%
where $\varphi_{i}^{(k)}(\lambda _{i})$ and $T_n^{(k)}(\lambda _{i})$ are
defined by Eq.~(\ref{tderivative}) with $\lambda_1\to \lambda_i\,
(k=0,1,2,...)$, and $\varepsilon $ is a small parameter.

It follows from Eq.~(\ref{kpg}) and
\begin{equation}
\lim\limits_{\varepsilon \rightarrow 0}\dfrac{T_n(\lambda _{i}+\varepsilon
)\varphi _{i}(\lambda _{i}+\varepsilon )}{\varepsilon ^{k_{i}}}=0,
\label{jixiankp}
\end{equation}%
with $i=1,2,...,M$ and $k_{i}=0,1,...,m_{i}$, that we obtain a linear
algebraic system with the $2N$ equations ($N=M+\sum_{i=1}^{M}m_{i}$):
\begin{equation}
\begin{array}{r}
T_n^{(0)}(\lambda _{i})\varphi _{i}^{(0)}(\lambda _{i})=0,\vspace{0.1in} \\
T_n^{(0)}(\lambda _{i})\varphi _{i}^{(1)}(\lambda _{i})+T_n^{(1)}(\lambda
_{i})\varphi _{i}^{(0)}(\lambda _{i})=0,\vspace{0.1in} \\
T_n^{(0)}(\lambda _{i})\varphi_i^{(2)}(\lambda _{i})+T_n^{(1)}(\lambda
_{1})\varphi_i^{(1)}(\lambda _{i})+T_n^{(2)}(\lambda
_{i})\varphi_i^{(0)}(\lambda _{i})=0,\vspace{0.1in} \\
\qquad \cdots \cdots ,\qquad \qquad \vspace{0.1in} \\
\sum\limits_{j=0}^{m_{i}}T_n^{(j)}(\lambda _{i})\varphi
_{i}^{(m_{i}-j)}(\lambda _{i})=0,%
\end{array}
\label{kpsysg}
\end{equation}%
in which some equations for index $i$, i.e., $T_n^{(0)}(\lambda _{i})\varphi
_{i}^{(0)}(\lambda _{i})=T_n(\lambda _{i})\varphi _{i}(\lambda _{i})=0$, are
similar to the first equations in~Eqs. (\ref{nls1sys}), but they are
different if there exists at least one $m_{i}\not=0$. When eigenvalues $%
\lambda _{i}$ $(i=1,2,..,M)$ are suitably chosen, so that the determinant of
the coefficients for system~(\ref{nls1sys}) does not vanish, the
transformation matrix $T$ is uniquely determined by system~(\ref{nls1sys}),
and Theorem 1 still holds. Using new functions $a_n^{(j)}$ and $b_n^{(j)}$
obtained with the $N$-order Darboux matrix $T_n$, we can derive the new DT
with $M$ distinct eigenvalues. Here, we identify Eqs.~(\ref{dt}) and (\ref%
{gauge}) as a generalized $(M,N-M)$-fold DTs for Eq.~(\ref{nls}), which is
specified by the following theorem.\newline
\newline
\textbf{Theorem 3.} \textit{Let $\varphi _{i}(\lambda _{i})=(\phi
_{i}(\lambda _{i}),\,\psi _{i}(\lambda _{i}))^{\mathrm{T}}$ be column-vector
solutions with distinct spectral parameters $\lambda _{i}\,(i=1,2,...,M)$ of
the problem based on Eqs. (\ref{lax1}) and (\ref{lax2}), and $%
(R_{0}(n,t),\,S_{0}(n,t))$ is the same seed solution of Eq.~(\ref{nls}), as
considered above, then the generalized $(M,N-M)$-fold DT for Eq.~(\ref{nls})
is given by
\begin{equation}
\widetilde{R}_{N}(n,t)=\dfrac{R_{0}(n,t)+\sigma {b_{n}^{(N-1)}}^{\ast }}{%
a_{n}^{(0)}},\ \ \ \widetilde{S}_{N}(n,t)=\sigma {%
b_{n+1}^{(0)}+S_{0}(n,t)a_{n+1}^{(0)}}~,  \label{nls1sol}
\end{equation}%
with $a_{n}^{(0)}=\dfrac{\Delta a_{n}^{(0)}}{\Delta _{N}^{\epsilon (M)}}%
,b_{n}^{(0)}=\dfrac{\Delta b_{n}^{(0)}}{\Delta _{N}^{\epsilon (M)}}%
,b_{n}^{(N-1)}=\dfrac{\Delta b_{n}^{(N-1)}}{\Delta _{N}^{\epsilon (M)}},\
\Delta _{N}^{\epsilon (M)}=\mathrm{det}([\Delta ^{(1)}...\Delta ^{(M)}]^{%
\mathrm{T}})$,\newline
where $\Delta ^{(i)}=(\Delta _{j,s}^{(i)})_{2(m_{i}+1)\times 2N},$\, in
which $\Delta _{j,s}^{(i)}\,\,(1\leq j\leq 2(m_{i}+1)$,\thinspace\ $1\leq
s\leq N$, $i=1,2,...,M$) are given by the following expressions:
\begin{equation}
\Delta _{j,s}^{(i)}=%
\begin{cases}
\sum\limits_{k=0}^{j-1}C_{N-s}^{k}{\lambda _{i}}^{(N-s-k)}{\phi ^{(j-1-k)}}
\quad \mathrm{for}\quad 1\leq j\leq m_{i}+1,\,1\leq s\leq N,\vspace{0.1in}
\\
\sum\limits_{k=0}^{j-1}C_{2N-s}^{k}{\lambda _{i}}^{(2N-s-k)}{\psi ^{(j-1-k)}}
\quad \mathrm{for}\quad 1\leq j\leq m_{i}+1,\,N+1\leq s\leq 2N,\vspace{0.1in}
\\
\sum\limits_{k=0}^{j-(N+1)}C_{s}^{k}{\lambda _{i}}^{\ast (s-k)}{\psi
^{*(j-N-1-k)}}\quad \mathrm{for}\quad m_{i}+2\leq j\leq 2(m_{i}+1),\,1\leq
s\leq N,\vspace{0.1in} \\
-\sigma \sum\limits_{k=0}^{j-(N+1)}C_{s-N}^{k}{\lambda _{i}}^{\ast (s-N-k)}{%
\phi ^{*(j-N-1-k)}}\quad \mathrm{for}\quad m_{i}+2\leq j\leq
2(m_{i}+1),\,N+1\leq s\leq 2N,
\end{cases}
\notag  \label{mrw2}
\end{equation}
with $\Delta a_{n}^{(0)}$, $\Delta b_{n}^{(0)}$ and $\Delta b_{n}^{(N-1)}$ given by determinant $\Delta _{N}^{\epsilon (M)}$, replacing its $N$-th, $(2N)$-th and $(N+1)$-th columns by the column vector $b=(b_{j})_{2N\times 1}$, with
\begin{equation}
b_{j}=
\begin{cases}
-\sum\limits_{k=0}^{j-1}C_{N}^{k}{\lambda _{i}}^{(N-k)}{\phi ^{(j-1-k)}}%
\quad 1\leq j\leq m_{i}+1,\vspace{0.1in} \\
-{\psi ^{*(j-N-1)}}\quad \mathrm{for}\quad m_{i}+2\leq j\leq 2(m_{i}+1).%
\end{cases}
\notag  \label{d3}
\end{equation}%
}

Notice that when $M=1$ and $m_{1}=N-1$, Theorem 3 reduces to Theorem 2, and
when $M=N$ and $m_{i}=0,\,1\leq i\leq N$, Theorem 3 reduces to Theorem 1.
When $M\neq 1$ and $M\neq $ $N$, the procedure makes it possible to derive
new DTs, which we do not explicitly discuss here. Below, we use the
generalized $(1,N-1)$-fold Darboux transformation to construct multi-RW
solutions of Eq.~(\ref{nls}) from the seed CW states.

\section{Higher-order vector rogue-wave solutions and dynamical behavior}

In what follows we produce multi-RW solutions in terms of the determinants
introduced above for Eq.~(\ref{nls}) with $\sigma =1$ (the self-focusing
nonlinearity) by means of the generalized $(1,N-1)$-fold DT. Equation~(\ref%
{nls}) is similar to the NLS equation, which gives rise to RWs in the case
of $\sigma =1$ (and it does not generate such solutions in the case of the
self-defocusing, with $\sigma =-1$). We here take, as the seed, CW solution (%
\ref{s1}) of Eq. (\ref{nls}) with $\sigma =1$. The substitution of this into
the Lax-pair equations (\ref{lax1}) and (\ref{lax2}) yields the following
solution:
\begin{equation}
\varphi (\lambda )=C_{1}\tau ^{n}e^{\rho t+\Theta (\varepsilon )}\left[
\begin{array}{c}
i\dfrac{1-\lambda -\sqrt{(\lambda -1)^{2}-4\lambda c^{2}}}{2\lambda c}%
\vspace{0.1in} \\
e^{2i(1+c^{2})t}%
\end{array}%
\right] ,  \label{phi}
\end{equation}%
with
\begin{equation}
\begin{array}{l}
\tau =\dfrac{\lambda ^{2}-2c^{2}\lambda +1+(\lambda +1)\sqrt{(\lambda
-1)^{2}-4\lambda c^{2}}}{2\lambda },\vspace{0.1in} \\
\rho =-\dfrac{\lambda ^{2}+2\lambda (c^{2}+1)-1+(\lambda -1)\sqrt{(\lambda
-1)^{2}-4\lambda c^{2}}}{2\lambda },\vspace{0.1in} \\
\Theta (\varepsilon )=\sqrt{(\lambda -1)^{2}-4\lambda c^{2}}%
\sum\limits_{k=1}^{N}(e_{k}+id_{k})\varepsilon ^{2k},%
\end{array}
\notag
\end{equation}%
where $C_{1}$ is an arbitrary constant, $e_{k},d_{k}$ $(k=1,2,...,N)$ are
free real coefficients, and $\varepsilon $ is an artificially introduced
small parameter.

Next, we fix the spectral parameter in Eq.~(\ref{phi}) as $\lambda =\lambda
_{1}+\varepsilon ^{2}$ with $\lambda _{1}=2c^{2}+1+2\sqrt{c^{2}(c^{2}+1)}$,
and expand vector function $\varphi $ in Eq.~(\ref{phi}) into the Taylor
series around $\varepsilon =0$, explicitly calculating the two first
coefficients of the expansion. In particular, choosing $C_{1}=1,c=\frac{3}{4}
$, which yields $\lambda _{1}=4$, we obtain
\begin{equation}
\varphi (\varepsilon ^{2})=\sum_{j=0}^{\infty}\varphi
^{(j)}\varepsilon^{2j}=\varphi ^{(0)}+\varphi ^{(1)}\varepsilon ^{2}+\varphi
^{(2)}\varepsilon ^{4}+\varphi ^{(3)}\varepsilon ^{6}+\cdots ,  \label{e271}
\end{equation}%
where
\begin{equation}
\varphi ^{(0)}=\left(
\begin{array}{c}
\phi ^{(0)} \\
\psi ^{(0)}%
\end{array}%
\right) =\frac{1}{2}\left( \frac{25}{16}\right) ^{n}\left(
\begin{array}{c}
ie^{-\frac{55}{16}it}\vspace{0.1in} \\
2e^{-\frac{5}{16}it}%
\end{array}%
\right) ,\qquad\qquad\qquad \hspace{0.3in}  \notag
\end{equation}%
\begin{equation}
\varphi ^{(1)}=\frac{1}{15360}\left( \frac{25}{16}\right) ^{n}\left[
\begin{array}{c}
e^{-\frac{55}{16}it}(7680t-2025it^{2}+2304in^{2}+4320nt+640i+3840in)\vspace{%
0.1in} \\
-6e^{-\frac{5}{16}it}(1360it+675t^{2}-768n^{2}+1440int)%
\end{array}%
\right] ,  \notag
\end{equation}%
$\varphi ^{(2,3)}=(\phi ^{(2,3)},\psi ^{(2,3)})^{\mathrm{T}}\,$are listed in
\textbf{Appendix A}, and $n$ denotes the discrete spatial variable.

It follows from Eqs.~(\ref{gauge}), (\ref{dt}), and (\ref{e271}) that we can
derive new exact RW solutions of Eq.~(\ref{nls}) as follows:
\begin{equation}
\hspace{1cm}\widetilde{R}_{N}(n,t)=\dfrac{R_{0}(n,t)+{b_{n}^{(N-1)}}^{\ast }%
}{a_{n}^{(0)}},\ \ \ \widetilde{S}_{N}(n,t)={%
b_{n+1}^{(0)}+S_{0}(n,t)a_{n+1}^{(0)}}.  \label{solu}
\end{equation}%
where $N$ denotes the number of iterations of the generalized perturbation
Darboux transformation. While the solution with $N=1$ reduces to the CW
seed, nontrivial solutions (\ref{solu}) of Eq.~(\ref{nls}) are analyzed
below for $N=2,3,4$.

\subsection{First-order vector rogue waves}

\textit{The structure of the first-order RW solutions}---. When $N=2$,
according to Theorem 2, we obtain the first-order vector RW solution of Eq.~(\ref%
{nls}) as
\begin{equation}
\begin{array}{ll}
\widetilde{R}_{2}(n,t)=\dfrac{3}{4}\dfrac{%
3600t+i(2025t^{2}+2304n^{2}+3072n-320)}{2025t^{2}+2304n^{2}+3072n+1280}e^{%
\frac{25}{8}it},\vspace{0.1in} &  \\
\widetilde{S}_{2}(n,t)=\dfrac{3}{4}\dfrac{%
-3600t+i(2025t^{2}+2304n^{2}+5376n+1792)}{2025t^{2}+2304n^{2}+5376n+3392}e^{-%
\frac{25}{8}it}, &
\end{array}
\label{rw-11}
\end{equation}%
which has no singularities. To understand the structure of this RW solution,
one can formally treat the discrete spatial variable $n$ as a continuous one, and first look at the
intensities:

%%%%%%%%%%%%%%%%%%%%%%%%%%%%%%%%%%%%%%%%%%%%%%%%%%%%
\begin{figure}[tbp]
\vspace{0.1in}
\par
\begin{center}
{\scalebox{0.6}[0.7]{\includegraphics{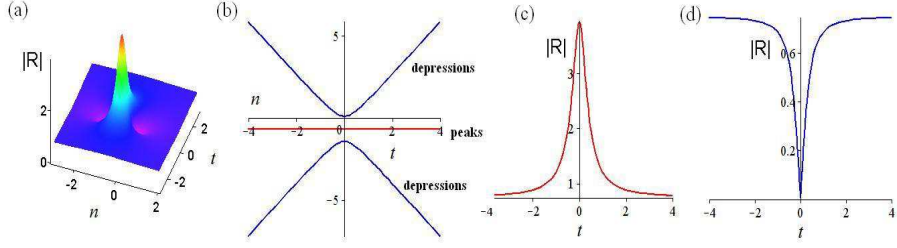}}}
\end{center}
\par
\vspace{-0.2in}
\caption{{\protect\small (color online). (a) The first-order rogue wave
solution $\widetilde{R}_{2}(n,t)$ given by Eq. (\protect\ref{rw-11}) with
one peak and two depression points. (b) The motion of the peak and
depression centers, as per Eqs.~(\protect\ref{th}) and (\protect\ref{tc}).
Other panels display the evolution of the absolute value of the field at (c)
the quiescent peak ($n=-\frac{2}{3}$) and (d) the moving depression points ($%
n=-\frac{2}{3}\pm \frac{1}{48}\protect\sqrt{1344+6075t^{2}}$).}}
\label{fig-rw1r}
\end{figure}
%%%%%%%%%%%%%%%%%%%%%%%%%%%%%%%%%%%%%%%%%%%%%%%%%%%%

%%%%%%%%%%%%%%%%%%%%%%%%%%%%%%%%%%%%%%%%%%%%%%%%%%%%
\begin{figure}[tbp]
\vspace{0.1in}
\par
\begin{center}
{\scalebox{0.6}[0.7]{\includegraphics{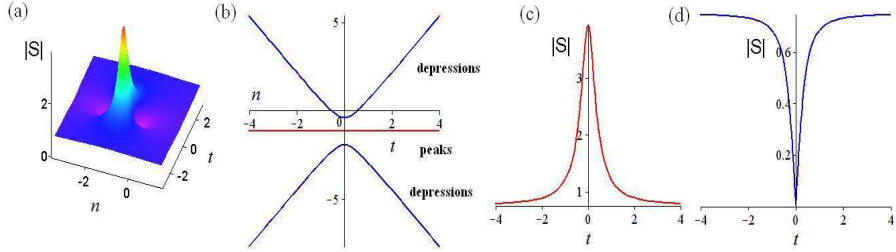}}}
\end{center}
\par
\vspace{-0.2in}
\caption{{\protect\small (color online). (a) The first-order rogue wave
solution $\widetilde{S}_{2}(n,t)$ given by Eq. (\protect\ref{rw-11}) with
one peak and two depression points. (b) The motion of the peak and
depression centers as per Eqs.~(\protect\ref{th1}) and (\protect\ref{tc1}).
Other panels display the evolution of the absolute value of the field at (c)
the quiescent peak ($n=-\frac{7}{6}$) and (d) the moving depression points ($%
n=-\frac{7}{6}\pm \frac{1}{48}\protect\sqrt{1344+6075t^{2}}$).}}
\label{fig-rw1s}
\end{figure}
%%%%%%%%%%%%%%%%%%%%%%%%%%%%%%%%%%%%%%%%%%%%%%%%%%%%

\begin{equation}
\begin{array}{ll}
|\widetilde{R}_{2}(n,t)|^{2}=\dfrac{9}{16}\dfrac{%
(2025t^{2}+2304n^{2}+3072n-320)^{2}+3600^{2}t^{2}}{%
(2025t^{2}+2304n^{2}+3072n+1280)^{2}}, \vspace{0.1in} &  \\
|\widetilde{S}_{2}(n,t)|^{2}=\dfrac{9}{16}\dfrac{%
(2025t^{2}+2304n^{2}+5376n+1792)^{2}+3600^{2}t^{2}}{%
(2025t^{2}+2304n^{2}+5376n+3392)^{2}}, &
\end{array}
\label{rw-12}
\end{equation}%
As follows from here, $|\widetilde{R}_{2}(n,t)|^{2}$ has three critical
points: $(n_{1},t_{1})=(-\frac{2}{3},0),(n_{2,3},t_{2,3})=(-\frac{4\pm \sqrt{%
21}}{6},0).$ The first point, $(n_{1},t_{1})$, is a maximum (peak), at which
$|\widetilde{R}_{2,\max }(n,t)|=\frac{63}{16}$. The two other critical
points $(n_{2,3},t_{2,3})$ are minima (depressions), at which the amplitude
vanishes. Further, the (modulationally unstable) background value is given
by $|\widetilde{R}_{2}(n,t)|\longrightarrow |\widetilde{R}_{2,\infty }(n,t)|=%
\frac{3}{4}$ at $|n|,t\longrightarrow \infty $. Thus, we find a relation
between the peak and background values: $|\widetilde{R}_{2,\max }(n,t)|=%
\frac{21}{4}|\widetilde{R}_{2,\infty }(n,t)|$ (see Fig.~\ref{fig-rw1r}(a)).
Moreover, we have the conservation law $\int\limits_{-\infty }^{+\infty }(|%
\widetilde{R}_{2}(n,t)|^{2}-\frac{9}{16})dn=0$, which follows from the fact
that the RW is actually formed from the flat background.

Next, intensity $|\widetilde{S}_{2}(n,t)|^{2}$ has three critical points: $%
(n_{1},t_{1})=(-\frac{7}{6},0),(n_{2,3},t_{2,3})=(-\frac{7\pm \sqrt{21}}{6}%
,0),$ the first again being a maximum point (peak), with the corresponding
maximum amplitude $|\widetilde{S}_{2,\max }(n,t)|=\frac{63}{16}$. Two other
critical points, $(n_{2,3},t_{2,3})$, are minima (depressions), with zero
amplitude. In this component, the background value is $|\widetilde{S}%
_{2}(n,t)|\longrightarrow |\widetilde{S}_{2,\infty }(n,t)|=\frac{3}{4}$ at $%
|n|,t\longrightarrow \infty $. Therefore, we have, as in the $\widetilde{R}%
_{2}$ component, $|\widetilde{S}_{2,\max }(n,t)|=\frac{21}{4}|\widetilde{S}%
_{2,\infty }(n,t)|$ (see Fig.~\ref{fig-rw1s}(a)), and the respective
conservation law, $\int\limits_{-\infty }^{+\infty }(|\widetilde{S}%
_{2}(n,t)|^{2}-\frac{9}{16})dn=0$.

For intensity $|\widetilde{R}_{2}(n,t)|^{2}$ given by Eq.~(\ref{rw-12}), the
trajectory of the peak's centers is produced by the dependence of its
spatial coordinate, $T_{rh}$, on $t$. It follows from the expression for $|%
\widetilde{R}_{2}(n,t)|^{2}$ that this coordinate actually stays constant,
\begin{equation}
\begin{array}{ll}
T_{rh}=-\dfrac{2}{3}, &
\end{array}
\label{th}
\end{equation}%
while the coordinates, $T_{rc\pm }$, of centers of the two depressions are
given by
\begin{equation}
\begin{array}{ll}
T_{rc\pm }=-\dfrac{2}{3}\pm \dfrac{1}{48}\sqrt{6075t^{2}+1344}. &
\end{array}
\label{tc}
\end{equation}%
The comparison of Eqs. (\ref{th}) and (\ref{tc}) produces the RW width,
which is defined as the distance between the two depressions:
\begin{equation}
\begin{array}{ll}
T_{rd}=\dfrac{1}{24}\sqrt{6075t^{2}+1344}. &
\end{array}
\label{td}
\end{equation}%
Figure~\ref{fig-rw1r}(b) displays the trajectories of the peak's centers, $%
T_{rh}$, and centers of the two depressions, $T_{rc\pm }$. Figures~\ref%
{fig-rw1r}(c) and \ref{fig-rw1r}(d) display profiles of $|\widetilde{R}%
_{2}(n,t)|$ at the peak ($n=-\frac{2}{3}$, as given by Eq.~(\ref{th})) and
depressions ($n=-\frac{2}{3}\pm \frac{1}{48}\sqrt{6075t^{2}+1344}$, as given
by Eq. (\ref{tc})).

For intensity $|\widetilde{S}_{2}(n,t)|^{2}$ given by Eq.~(\ref{rw-12}), the
trajectory of the peak's centers is determined by the dependence of its
spatial coordinate, $T_{sh}$, on $t$. It follows from the expression for $|%
\widetilde{S}_{2}(n,t)|^{2}$ that
\begin{equation}
\begin{array}{ll}
T_{sh}=-\dfrac{7}{6}, &
\end{array}
\label{th1}
\end{equation}%
and the spatial coordinates, $T_{sc\pm }$, of centers of the two depressions
are given by
\begin{equation}
\begin{array}{ll}
T_{sc\pm }=-\dfrac{7}{6}\pm \dfrac{1}{48}\sqrt{6075t^{2}+1344}, &
\end{array}
\label{tc1}
\end{equation}%
leading to the RW width, which is again defined as the spatial distance
between two depressions:
\begin{equation}
\begin{array}{ll}
T_{sd}=\dfrac{1}{24}\sqrt{6075t^{2}+1344}, &
\end{array}
\label{td1}
\end{equation}%
Figure~\ref{fig-rw1s}(b) displays the trajectories of the peak's centers, $%
T_{sh}$, and of the centers of the two depressions, $T_{sc\pm }$. Figures~%
\ref{fig-rw1s}(c) and~\ref{fig-rw1s}(d) illustrate the profiles of $|%
\widetilde{R}_{2}(n,t)|$ at the peak ($n=-\frac{7}{6}$, as given by Eq.~(\ref%
{th1})) and depressions ($n=-\frac{7}{6}\pm \frac{1}{48}\sqrt{6075t^{2}+1344}
$, as given by Eq.~(\ref{tc1})).

\begin{figure}[tbp]
\begin{center}
{\scalebox{0.6}[0.6]{\includegraphics{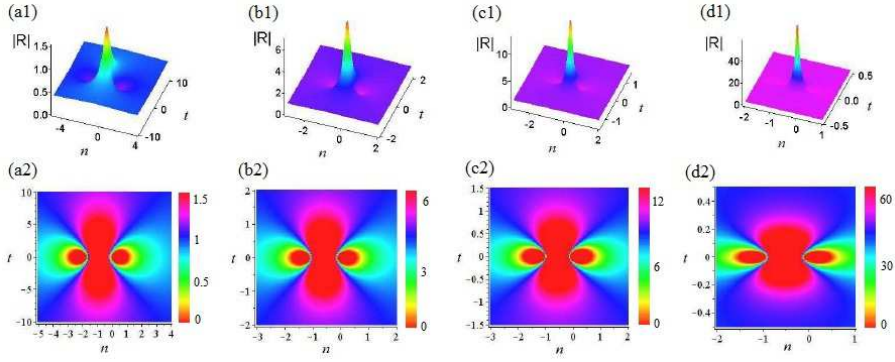}}}
\end{center}
\par
\vspace{-0.15in}
\caption{{\protect\small (color online). The first-order rogue-wave solution
$\widetilde{R}_{2}(n,t)$, given by Eq.~(\protect\ref{solu}) with $N=2$:
(a1)-(a2) $c=\frac{5}{12}$ (i.e., $\protect\lambda =\frac{9}{4}$); (b1)-(b2)
$c=1$ (i.e., $\protect\lambda =3+2\protect\sqrt{2}$); (c1)-(c2) $c=\frac{4}{3%
}$ (i.e., $\protect\lambda =9$); and (d1)-(d2) $c=\frac{12}{5}$ (i.e., $%
\protect\lambda =25$). In this figure and similar figures following below,
the bottom row shows the top view of the generated RWs (note the difference
in the scale of variable $t$ in comparison with the 3D plots displayed in
the top row). In other figures showing top views, the color-code bar is the same
as used here.}}
\label{fig-rw1bjr}
\end{figure}

\begin{figure}[tbp]
\begin{center}
{\scalebox{0.6}[0.6]{\includegraphics{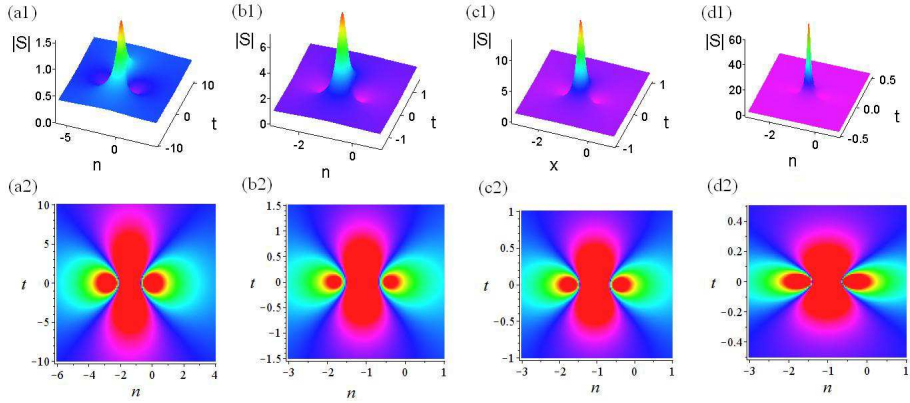}}}
\end{center}
\par
\vspace{-0.15in}
\caption{{\protect\small (color online). The first-order rogue wave solution
$\widetilde{S}_{2}(n,t)$ given by Eq.~(\protect\ref{solu}) with $N=2$:
(a1)-(a2) $c=\frac{5}{12}$ (i.e., $\protect\lambda =\frac{9}{4}$); (b1)-(b2)
$c=1$ (i.e., $\protect\lambda =3+2\protect\sqrt{2}$); (c1)-(c2) $c=\frac{4}{3%
}$ (i.e., $\protect\lambda =9$); and (d1)-(d2) $c=\frac{12}{5}$ (i.e., $%
\protect\lambda =25$).}}
\label{fig-rw1bjs}
\end{figure}

In addition, it is relevant to note that the amplitude $c$ of the CW background
can control the first-order RW solution (\ref{solu}) with $N=2$. Here we use
values of $c$ to modulate the shape, central position, and amplitude of the
first-order RW solution (see Figs.~\ref{fig-rw1bjr} and~\ref{fig-rw1bjs}).

\begin{itemize}
\item {} For $c=\frac{5}{12}$ (i.e., $\lambda =\frac{9}{4}$), $\widetilde{R}%
_{2}(n,t)$ peaks around the point $(n_1,t_1)=(-\frac{9}{10},0)$, and the
background value is determined by $|\widetilde{R}_{2}(n,t)|\longrightarrow |%
\widetilde{R}_{2,\infty }(n,t)|=\frac{5}{12}$ at $n,t\longrightarrow \infty $%
. The maximum value $|\widetilde{R}_{2,\max }(n,t)|$ of $|\widetilde{R}%
_{2}(n,t)|$ is $\frac{665}{432}$ at point $(n_1,t_1)=(-\frac{9}{10},0),$ and
the minimum is zero, at $(n_{2,3},t_{2,3})=(-\frac{9\pm \sqrt{133}}{10},0)$,
see Figs.~\ref{fig-rw1bjr}a1-a2;

\item {} When $c=1$ (i.e., $\lambda =3+2\sqrt{2}$), $\widetilde{R}_{2}(n,t)$
peaks around the point $(n_1,t_1)=(-\frac{1+\sqrt{2}}{4},0)$, and the
background corresponds to $|\widetilde{R}_{2}(n,t)|\longrightarrow |%
\widetilde{R}_{2,\infty }(n,t)|=1$ at $n,t\longrightarrow \infty $. The
maximum value $|\widetilde{R}_{2,\max }(n,t)|$ of $|\widetilde{R}_{2}(n,t)|$
is $7$ at $(n_1,t_1)=(-\frac{1+\sqrt{2}}{4},0)$, and the minimum is zero at $%
(n_{2,3},t_{2,3})=(-\frac{1+\sqrt{2}\pm \sqrt{7}}{4},0)$, see Figs.~\ref%
{fig-rw1bjr}(b1-b2);

\item {} For $c=\frac{4}{3}$ (i.e., $\lambda =9$), $\widetilde{R}_{2}(n,t)$
peaks around the point $(n_1,t_1)=(-\frac{9}{16},0)$, with the background
corresponding to $|\widetilde{R}_{2}(n,t)|\longrightarrow |\widetilde{R}%
_{2,\infty }(n,t)|=\frac{4}{3}$ at $n,t\longrightarrow \infty $. The maximum
value $|\widetilde{R}_{2,\max }(n,t)|$ of $|\widetilde{R}_{2}(n,t)|$ is $%
\frac{364}{27}$ at $(n_1,t_1)=(-\frac{9}{16},0)$ and the minimum value is
again zero, at $(n_{2,3},t_{2,3})=(-\frac{9\pm \sqrt{91}}{16},0)$, see Figs.~%
\ref{fig-rw1bjr}(c1-c2);

\item {} When $c=\frac{12}{5}$ (i.e., $\lambda =25$), $\widetilde{R}_{2}(n,t)
$ peaks around the point $(n_1,t_1)=(-\frac{25}{48},0)$, and the background
corresponds to $|\widetilde{R}_{2}(n,t)|\longrightarrow |\widetilde{R}%
_{2,\infty }(n,t)|=\frac{12}{5}$ at $n,t\longrightarrow \infty $. The
maximum value $|\widetilde{R}_{2,\max }(n,t)|$ of $|\widetilde{R}_{2}(n,t)|$
is $\frac{7812}{125}$ at $(n_1,t_1)=(-\frac{25}{48},0)$, and the minimum is,
once again, zero, at $(n_{2,3},t_{2,3})=(-\frac{25\pm \sqrt{651}}{48},0)$,
see Figs.~\ref{fig-rw1bjr}(d1-d2).
\end{itemize}

From the above analysis, we conclude that the increase of spectral parameter
$\lambda $ corresponds to the increase of $c$, and the amplitude of the
first-order RW grows too with the increase of $c$. The central position and
peak of the first-order RW move to the right along line $t=0$. A similarly
picture is produced by the consideration of the other component.

%%%%%%%%%%%%%%%%%%%%%%%%%%%%%%%%%%%%%%%%%%%%%%%%%%%%
%%%%%%%%%%%%%%%%%%%%%%%%%%%%%%%%%%%%%%%%%%%%%%%%%%%%
%%%%%%%%%%%%%%%%%%%%%%%%%%%%%%%%%%%%%%%%%%%%%%%%%%%%
%%%%%%%%%%%%%%%%%%%%%%%%%%%%%%%%%%%%%%%%%%%%%%%%%%%%

\textit{Dynamical behavior}---. To further study the wave propagation in the
framework of the above discrete RW solutions, we here compare the exact
solutions and their perturbed counterparts, produced by numerical
simulations of Eq.~(\ref{nls}) with the initial conditions given by these
solutions with small perturbations.

\begin{figure}[tbp]
\begin{center}
{\scalebox{0.56}[0.65]{\includegraphics{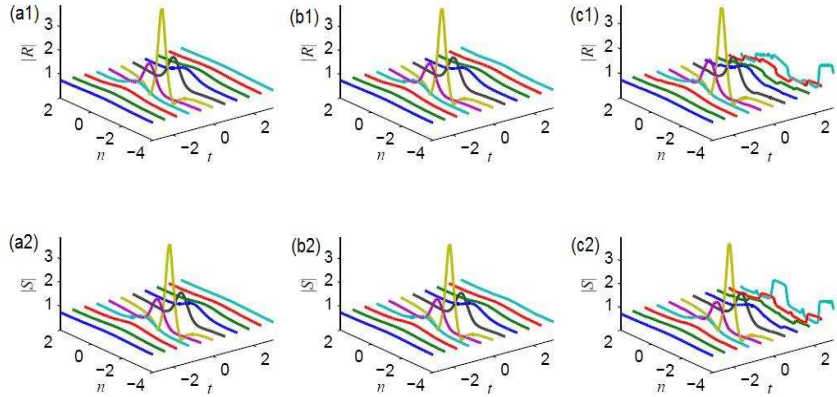}}}
\end{center}
\par
\vspace{-0.25in}
\caption{{\protect\small (color online). First-order RW solutions (\protect
\ref{rw-12}). The exact solution (left); the evolution generated by
simulations using exact solutions (\protect\ref{rw-12}) as an initial
conditions (middle); and the evolution generated by adding random noise with
amplitude }${\protect\small 0.02}${\protect\small \ to the initial
conditions (right).}}
\label{fig-rw1-noise}
\end{figure}

Figure~\ref{fig-rw1-noise} exhibits the exact first-order RW solution, $%
\widetilde{R}_{2}(n,t)$ and $\widetilde{S}_{2}(n,t)$, as given by Eq.~(\ref%
{rw-12}), and the one perturbed by small noise with amplitude $0.02$.
Figures~4(a1)-(a2) and (b1)-(b2) show that the time evolution of the RW
without the addition of the perturbation is very close to the corresponding
exact RW solution (\ref{rw-12}) in a short time interval, $t\in (-2.5,2.5)$.
However, if the small random noise is added to the initial condition (i.e.,
the RW solution (\ref{rw-12}) taken at $t=-2.5$), then the evolution
exhibits weak perturbations at $t<2.5$, and conspicuous perturbations later,
see Figs.~\ref{fig-rw1-noise}(c1-c2).
%%%%%%%%%%%%%%%%%%%%%%%%%%%%%%%%%%%%%%%%%%%%%%%%%%%%
%%%%%%%%%%%%%%%%%%%%%%%%%%%%%%%%%%%%%%%%%%%%%%%%%%%%

\subsection{Second-order vector rogue waves}

\textit{The structure of the second-order RWs}---. Theorem 2 with $N=3$
produces the second-order vector RW solution of Eq. (\ref{nls})
\begin{equation}
\widetilde{R}_{3}(n,t)=\dfrac{R_{0}(n,t)+{b_{n}^{(2)}}^{\ast }}{a_{n}^{(0)}}=%
\dfrac{3A_{1}(n,t)}{4B_{1}(n,t)}e^{\frac{25}{8}it},\, \, \, \widetilde{S}%
_{3}(n,t)={b_{n+1}^{(0)}+S_{0}(n,t)a_{n+1}^{(0)}}=\dfrac{3A_{2}(n,t)}{%
4B_{2}(n,t)}e^{-\frac{25}{8}it},  \label{rw2}
\end{equation}%
where $A_{j},B_{j}\,(j=1,2)$ are given by the following complex expressions:
\begin{equation}
\begin{array}{rl}
A_{1}(n,t)= &
16986931200tn^{3}+95551488000itd_{1}n^{2}-74649600000t^{2}d_{1}+59719680000it^{2}e_{1}
\\
&
-96259276800ine_{1}-4039114752in^{2}+3149280000it^{4}n^{2}-24286003200it^{2}n
\\
&
-67947724800in^{2}e_{1}-33973862400in^{3}e_{1}-27993600000it^{3}d_{1}+212336640000ie_{1}^{2}
\\
&
+3583180800it^{2}n^{4}-11346739200in^{2}t^{2}+53084160000itd_{1}+4199040000it^{4}n
\\
& +9555148800it^{2}n^{3}+4920750000t^{5}+1358954496in^{6}+150994944in^{3} \\
& +127401984000itnd_{1}+5435817984in^{5}+212336640000id_{1}^{2}+26214400i \\
& +922640625it^{6}+14929920000nt^{3}-11943936000it^{2}-6488064000t \\
& -2390753280in+6370099200n^{2}t+47185920000d_{1}+6220800000t^{3} \\
&
+84934656000n^{2}d_{1}+11197440000n^{2}t^{3}+6370099200n^{4}t-4811400000it^{4}
\\
& +6002049024in^{4}+113246208000nd_{1}+106168320000te_{1}+159252480000nte_{1}
\\
& -7785676800nt-47185920000ie_{1}+89579520000it^{2}ne_{1},%
\end{array}
\notag
\end{equation}%
\begin{equation}
\begin{array}{rl}
B_{1}(n,t)= &
-25480396800ne_{1}-4379443200nt^{2}+3583180800n^{2}t^{2}+5435817984n^{5} \\
& +1358954496n^{6}+2013265920n+127401984000ntd_{1}+9555148800n^{3}t^{2} \\
&
+3149280000n^{2}t^{4}+89579520000nt^{2}e_{1}+4199040000nt^{4}-33973862400n^{3}e_{1}
\\
&
+3583180800n^{4}t^{2}+95551488000n^{2}td_{1}+1534464000t^{2}+6123600000t^{4}
\\
&
-13271040000td_{1}+212336640000e_{1}^{2}+212336640000d_{1}^{2}+922640625t^{6}
\\
&
-27993600000t^{3}d_{1}+59719680000t^{2}e_{1}+4690280448n^{2}+7700742144n^{3}
\\
& +8833204224n^{4}-67947724800n^{2}e_{1}+419430400,%
\end{array}
\notag
\end{equation}%
\begin{equation}
\begin{array}{rl}
A_{2}(n,t)= &
-118908518400ne_{1}+8161689600nt^{2}+23290675200n^{2}t^{2}+9512681472n^{5}
\\
& +1358954496n^{6}+4304404480+18849202176n+222953472000ntd_{1} \\
&
+16721510400n^{3}t^{2}+3149280000n^{2}t^{4}+89579520000nt^{2}e_{1}+7348320000nt^{4}
\\
&
-33973862400n^{3}e_{1}-33973862400e_{1}+3583180800n^{4}t^{2}+95551488000n^{2}td_{1}
\\
& +1658880000t^{2}+9010440000t^{4}+74317824000td_{1}+212336640000e_{1}^{2}
\\
&
+212336640000d_{1}^{2}+922640625t^{6}-27993600000t^{3}d_{1}+104509440000t^{2}e_{1}
\\
& +37559992320n^{2}+42354081792n^{3}+27518828544n^{4}-118908518400n^{2}e_{1}.
\\
&
\end{array}%
\qquad \quad  \notag
\end{equation}%
\begin{equation}
\begin{array}{rl}
B_{2}(n,t)= &
-29727129600tn^{3}+104509440000it^{2}e_{1}+222953472000itnd_{1}+89579520000it^{2}ne_{1}
\\
&
-26674790400it^{2}n-25505280000it^{2}+95551488000itd_{1}n^{2}+74649600000t^{2}d_{1}
\\
& +24687673344in^{4}-189687398400ine_{1}-1362100224in+140673024000itd_{1} \\
& -118908518400in^{2}e_{1}+16721510400it^{2}n^{3}+3583180800it^{2}n^{4} \\
&
-33973862400in^{3}e_{1}+13259243520in^{2}+7348320000it^{4}n+3149280000it^{4}n^{2}
\\
&
+9512681472in^{5}+29142024192in^{3}-116549222400ie_{1}-27993600000it^{3}d_{1}
\\
&
-1924560000it^{4}-4920750000t^{5}+1358954496in^{6}+922640625it^{6}-1593835520i
\\
& +8360755200in^{2}t^{2}-26127360000nt^{3}+212336640000ie_{1}^{2}+6266880000t
\\
&
-41405644800n^{2}t-125042688000d_{1}-16485120000t^{3}+212336640000id_{1}^{2}
\\
&
-84934656000n^{2}d_{1}-11197440000n^{2}t^{3}-6370099200n^{4}t-198180864000nd_{1}
\\
& -185794560000te_{1}-159252480000nte_{1}-14509670400nt,%
\end{array}
\notag
\end{equation}

\begin{figure}[!t]
\begin{center}
{\scalebox{0.6}[0.6]{\includegraphics{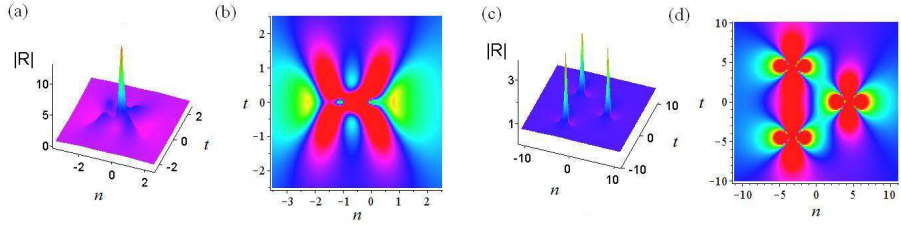}}}
\end{center}
\par
\vspace{-0.15in}
\caption{{\protect\small (color online). The second-order RW
solution $\widetilde{R}_{3}(n,t)$ given by Eq.~(\protect\ref{rw2}): (a)-(b) $%
e_{1}=d_{1}=0$; (c)-(d) -- a triangular pattern with $e_{1}=10,d_{1}=0$.}}
\label{fig-rw2-r}
\end{figure}

\begin{figure}[tbp]
\begin{center}
{\scalebox{0.6}[0.6]{\includegraphics{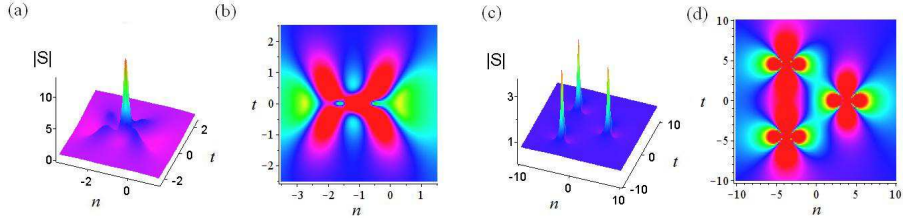}}}
\end{center}
\par
\vspace{-0.15in}
\caption{{\protect\small (color online). The second-order RW
solution $\widetilde{S}_{3}(n,t)$ given by Eq.~(\protect\ref{rw2}): (a)-(b) $%
e_{1}=d_{1}=0$; (c)-(d) -- a triangular pattern with $e_{1}=10,d_{1}=0$.}}
\label{fig-rw2-s}
\end{figure}

The profiles of the second-order RW solutions (\ref{rw2}) are displayed in
Figs.~\ref{fig-rw2-r} and~\ref{fig-rw2-s}. The pair of introduced parameters
$e_{1}$ and $d_{1}$ can be used to control strong and weak interactions of
the second-order RW solutions (\ref{rw2}), by which we mean, respectively,
tightly and loosely bound patterns:

\begin{itemize}
\item {} For $e_{1}=d_{1}=0$, we observe the strong interaction, see Figs.~\ref%
{fig-rw2-r}(a,b) and Figs.~\ref{fig-rw2-s}(a,b), which exhibit five local
maxima (peaks) and four minima (holes).

\item {} For $e_{1}d_{1}\neq 0$, we observe the weak interaction, see Figs.~\ref%
{fig-rw2-r}(c)-(d) and Figs.~\ref{fig-rw2-s}(c)-(d), in which the
second-order RW is split into three first-order RWs, whose centers form a
rotating triangle, and there are three local maxima (peaks) and six minima
(holes) exhibited by every profile. Moreover, we find that the size of the
triangle increase with the increase of $|e_{1}|$ or $|d_{1}|$, and $d_{1}$
controls the rotation of the triangle profile. When both $|e_{1}|$ and $%
|d_{1}|$ drop to zero, the three split first-order RWs feature strong
interaction (cf. Figs.~\ref{fig-rw2-r}(c)-(d) or \ref{fig-rw2-s}(c)-(d) and
Figs.~\ref{fig-rw2-r}(a)-(b) or \ref{fig-rw2-s}(a)-(b)), and the local
maxima grow, while the minimum values decrease, which is an effect of the
strong interaction between the first-order RWs. We do not report here the
dependence of the angular velocity of the rotation on the system's
parameters, as this analysis requires extremely heavy simulations.
\end{itemize}

%%%%%%%%%%%%%%%%%%%%%%%%%%%%%%%%%%%%%%%%%%%%%%%%%%%%
%%%%%%%%%%%%%%%%%%%%%%%%%%%%%%%%%%%%%%%%%%%%%%%%%%%%
%%%%%%%%%%%%%%%%%%%%%%%%%%%%%%%%%%%%%%%%%%%%%%%%%%%%
%%%%%%%%%%%%%%%%%%%%%%%%%%%%%%%%%%%%%%%%%%%%%%%%%%%%

\begin{figure}[tbp]
\begin{center}
{\scalebox{0.56}[0.7]{\includegraphics{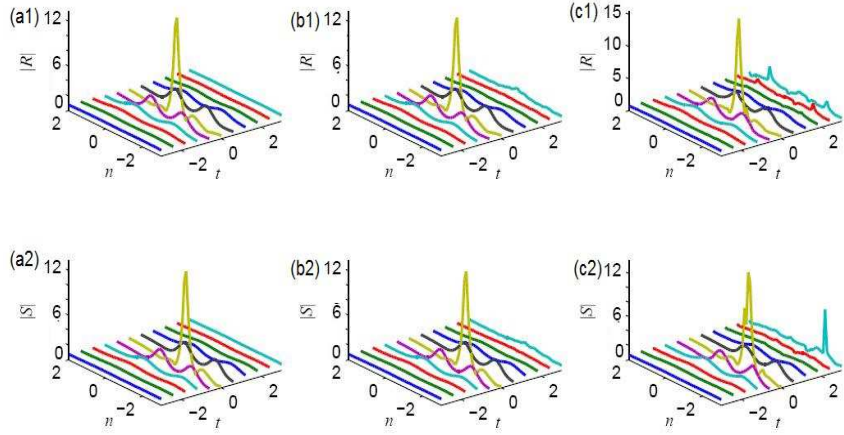}}}
\end{center}
\par
\vspace{-0.3in}
\caption{{\protect\small (color online). The second-order RW solutions (%
\protect\ref{rw2}) with $e_{1}=d_{1}=0$. The exact solution (left), the
simulated evolution using the exact solutions (\protect\ref{rw2}) as the
initial conditions (middle), and the evolution initiated by the exact
solution perturbed by relatively strong random noise with amplitude }$%
{\protect\small 0.05}${\protect\small .}}
\label{fig-rw2-noise}
\end{figure}

\textit{Dynamical behavior}---. Next, we study the dynamical behavior of the
second-order RWs, $\widetilde{R}_{3}(n,t)$ and $\widetilde{S}_{3}(n,t)$,
given by Eq.~(\ref{rw2}), by means of numerical simulations. We consider
cases of the strong (Figs.~\ref{fig-rw2-r}(a) and ~\ref{fig-rw2-s}(a)) and
weak (Figs.~\ref{fig-rw2-r}c and ~\ref{fig-rw2-s}c) interaction. Figures~\ref%
{fig-rw2-noise}(b1)-(b2) and (a1)-(a2) show that the evolution of the
second-order RW solution $\widetilde{R}_{3}(n,t)$ in the regime of the
strong interaction without the addition of noise almost exactly corroborates
the corresponding exact RW solution~(\ref{rw2}), in the time interval $t\in
(-2.5,\,2.5)$. If random noise with amplitude $0.05$ (which is larger than
the noise amplitude considered above) is added to the initial condition,
taken as per RW solution~(\ref{rw2}) with $e_{1}=d_{1}=0$ at $t=-2.5$, then
the evolution exhibits only relatively small perturbations, in comparison
with the unperturbed solution, see Figs.~\ref{fig-rw2-noise}(c1)-(c2).

\begin{figure}[tbp]
\begin{center}
{\scalebox{0.56}[0.67]{\includegraphics{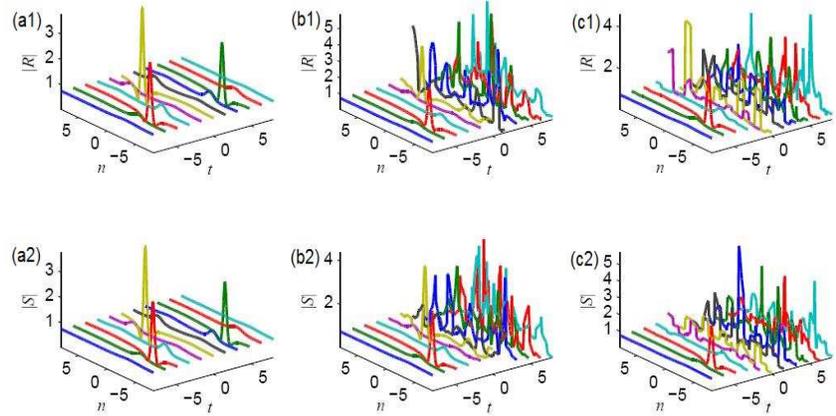}}}
\end{center}
\par
\vspace{-0.25in}
\caption{{\protect\small (color online). The second-order RW solutions (%
\protect\ref{rw2}) with $e_{1}=10,d_{1}=0$. The exact solution (left),
simulated evolution with the exact solution (\protect\ref{rw2}) used as the
initial conditions (middle), and the evolution initiated by the exact
solutions perturbed by weak random noise with amplitude }${\protect\small %
0.01}${\protect\small .}}
\label{fig-rw21-noise}
\end{figure}

However, in the case of the weak interaction (see Figs.~\ref{fig-rw2-r}(c)
and ~\ref{fig-rw2-s}(c)) the results are different. The second-order RW
solution~(\ref{rw2}) exhibits perturbations at $t>0$ even without adding the
noise to the initial conditions, which indicates strong instability of the
RW solution in this case, see Figs.~\ref{fig-rw21-noise}(b1)-(b2). If really
weak random noise , with amplitude $0.01$, is added to the initial solution,
i.e., to RW (\ref{rw2}) with $e_{1}=10,\, d_{1}=0$ at $t=-8$, the simulated
evolution exhibits obviously strong instability at $t>0$, see Figs.~\ref%
{fig-rw21-noise}(c1)-(c2).

Thus, Figs.~\ref{fig-rw2-noise}(c1)-(c2) and ~\ref{fig-rw21-noise}%
(c1)-(c2) demonstrate that the weak random noise produces a weak (strong)
effect in the case of the strong (weak) interaction. The difference may be
explained by the fact that, in the former case (strong interaction), the
energy is concentrated close to the origin in $(n,t)$ plane, see Figs.~\ref%
{fig-rw2-noise}(a1)-(a2), while in the latter case (weak interaction) case
the energy is spread around three points in the $(n,t)$ plane, see Figs.~\ref%
{fig-rw21-noise}(a1)-(a2), making this loosely bound pattern more amenable
to perturbation effects.%
%%%%%%%%%%%%%%%%%%%%%%%%%%%%%%%%%%%%%%%%%%%%%%%%%%%%
%%%%%%%%%%%%%%%%%%%%%%%%%%%%%%%%%%%%%%%%%%%%%%%%%%%%
%%%%%%%%%%%%%%%%%%%%%%%%%%%%%%%%%%%%%%%%%%%%%%%%%%%%
%%%%%%%%%%%%%%%%%%%%%%%%%%%%%%%%%%%%%%%%%%%%%%%%%%%%

\subsection{Third-order vector rogue waves}

\textit{The structure of the third-order RWs}---. Theorem 2 with $N=4$
produces the third-order vector RW solution of Eq. (\ref{nls}):
\begin{equation}
\widetilde{R}_{4}(n,t)=\dfrac{R_{0}(n,t)+{b_{n}^{(3)}}^{\ast }}{a_{n}^{(0)}}%
,\ \ \ \widetilde{S}_{4}(n,t)={b_{n+1}^{(0)}+S_{0}(n,t)a_{n+1}^{(0)}}.
\label{rw3}
\end{equation}%
An explicit analytical expression for this solution is not written here, as
it is very cumbersome. Variation of control parameters $e_{1},\,e_{2},%
\,d_{1},\,d_{2}$ casts the third-order RW into different forms.

\begin{figure}[tbp]
\begin{center}
{\scalebox{0.6}[0.6]{\includegraphics{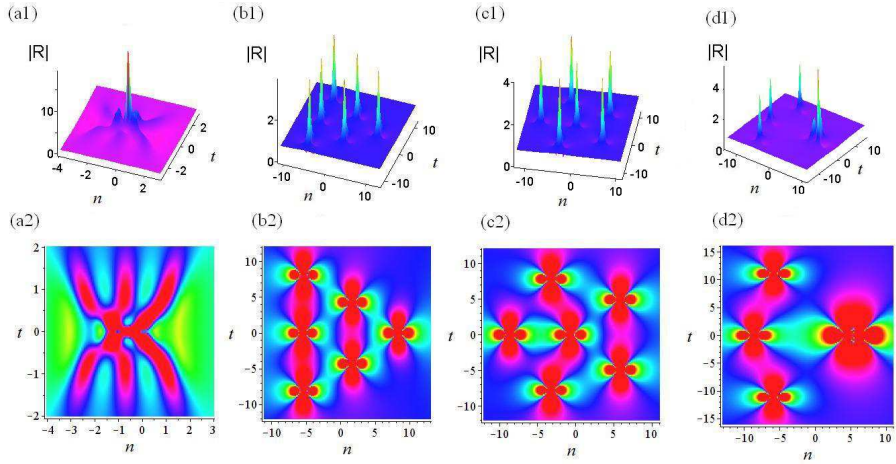}}}
\end{center}
\par
\vspace{-0.2in}
\caption{{\protect\small (color online). The third-order RW solution
$\widetilde{R}_{4}(n,t)$ given by Eq.~(\protect\ref{rw3}): (a1)-(a2) $%
e_{1,2}=d_{1,2}=0$; (b1)-(b2) -- a triangular pattern with $%
e_{1}=10,d_{1,2}=e_{2}=0$; (c1)-(c2) -- a pentagon pattern with $%
e_{2}=100,e_{1}=d_{1,2}=0$; and (d1)-(d2) display a quadrangle pattern
with $e_{1}=17.6875,e_{2}=230,d_{1,2}=0$.}}
\label{rw3-r}
\end{figure}

\begin{figure}[t]
\begin{center}
{\scalebox{0.6}[0.6]{\includegraphics{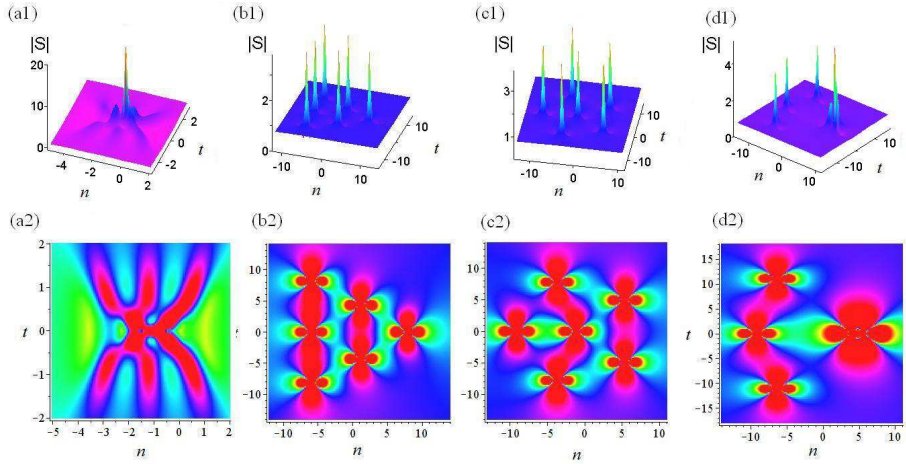}}}
\end{center}
\vspace{-0.25in}
\caption{{\protect\small (color online). The third-order RW solution
$\widetilde{S}_{4}(n,t)$ given by Eq.~(\protect\ref{rw3}): (a1)-(a2) $%
e_{1,2}=d_{1,2}=0$; (b1)-(b2) -- a triangular pattern with $%
e_{1}=10,d_{1,2}=e_{2}=0$; (c1)-(c2) -- a pentagon pattern with $%
e_{2}=100,e_{1}=d_{1,2}=0$; and (d1)-(d2) display a quadrangle pattern
with $e_{1}=17.6875,e_{2}=230,d_{1,2}=0$.}}
\label{rw3-s}
\end{figure}

\begin{itemize}
\item {} \thinspace\ At $e_{1,2}=d_{1,2}=0$, the third-order exists in the
regime of strong interaction, with the corresponding density graphs
displayed in Figs.~\ref{rw3-r}(a1)-(a2) and~\ref{rw3-s}(a1)-(a2).

\item {} \thinspace\ At $e_{1}=10,\,d_{1,2}=e_{2}=0$, the weak interaction
allows splitting of the third-order RW into six first-order RWs, which form
a triangular pattern, see Figs.~\ref{rw3-r}(b1)-(b2) and~\ref{rw3-s}%
(b1)-(b2).

\item {} \thinspace\ At $e_{2}=100,\,d_{1,2}=e_{2}=0$, the weak interaction
in the third-order RW also leads to splitting into six first-order RWs,
which form a rotating pentagon pattern with a first-order RW located near
the center, see Figs.~\ref{rw3-r}(c1)-(c2) and~\ref{rw3-s}(c1)-(c2).

\item {} \thinspace\ If we choose one nonzero parameter in each sets $\{e_{1},d_{1}\}$ and $\{e_{2},\,d_{2}\}$ (e.g., $e_{1}=17.6875,e_{2}=230,d_{1,2}=0$), then the third-order RW solution (\ref{rw3}) displays a different quadrangle pattern, that is, a second-order RW with the highest peak on the right side, and three first-order RWs that form an arc on the left side, see Figs.~\ref{rw3-r}(c1)-(c2) and~\ref{rw3-s}(c1)-(c2).
\end{itemize}

%%%%%%%%%%%%%%%%%%%%%%%%%%%%%%%%%%%%%%%%%%%%%%%%%%%%
%%%%%%%%%%%%%%%%%%%%%%%%%%%%%%%%%%%%%%%%%%%%%%%%%%%%
%%%%%%%%%%%%%%%%%%%%%%%%%%%%%%%%%%%%%%%%%%%%%%%%%%%%
%%%%%%%%%%%%%%%%%%%%%%%%%%%%%%%%%%%%%%%%%%%%%%%%%%%%

\begin{figure}[tbp]
\begin{center}
{\scalebox{0.56}[0.7]{\includegraphics{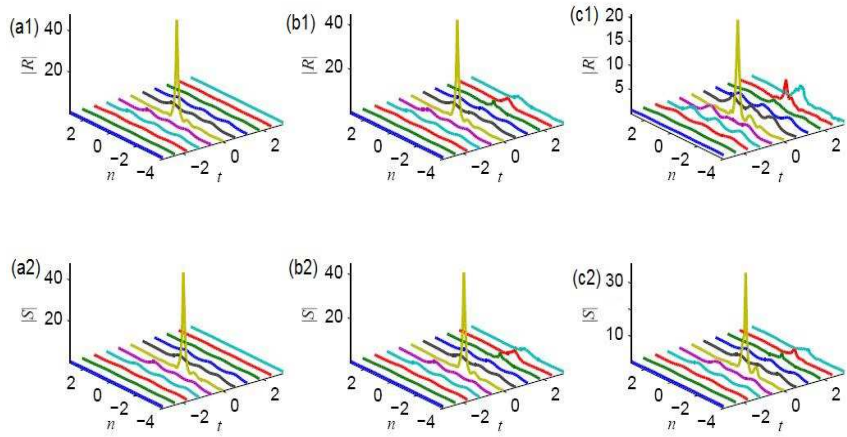}}}
\end{center}
\par
\vspace{-0.3in}
\caption{{\protect\small (color online). The third-order RW solutions (%
\protect\ref{rw3}) with $e_{1,2}=d_{1,2}=0$. The exact solution (left);
simulated evolution using exact solutions (\protect\ref{rw3}) as the initial
conditions (middle); and the evolution initiated by the exact solutions
perturbed by relatively strong random noise with amplitude }${\protect\small %
0.08}${\protect\small \ (right).}}
\label{fig-rw3-noise}
\end{figure}

\begin{figure}[tbp]
\begin{center}
{\scalebox{0.56}[0.7]{\includegraphics{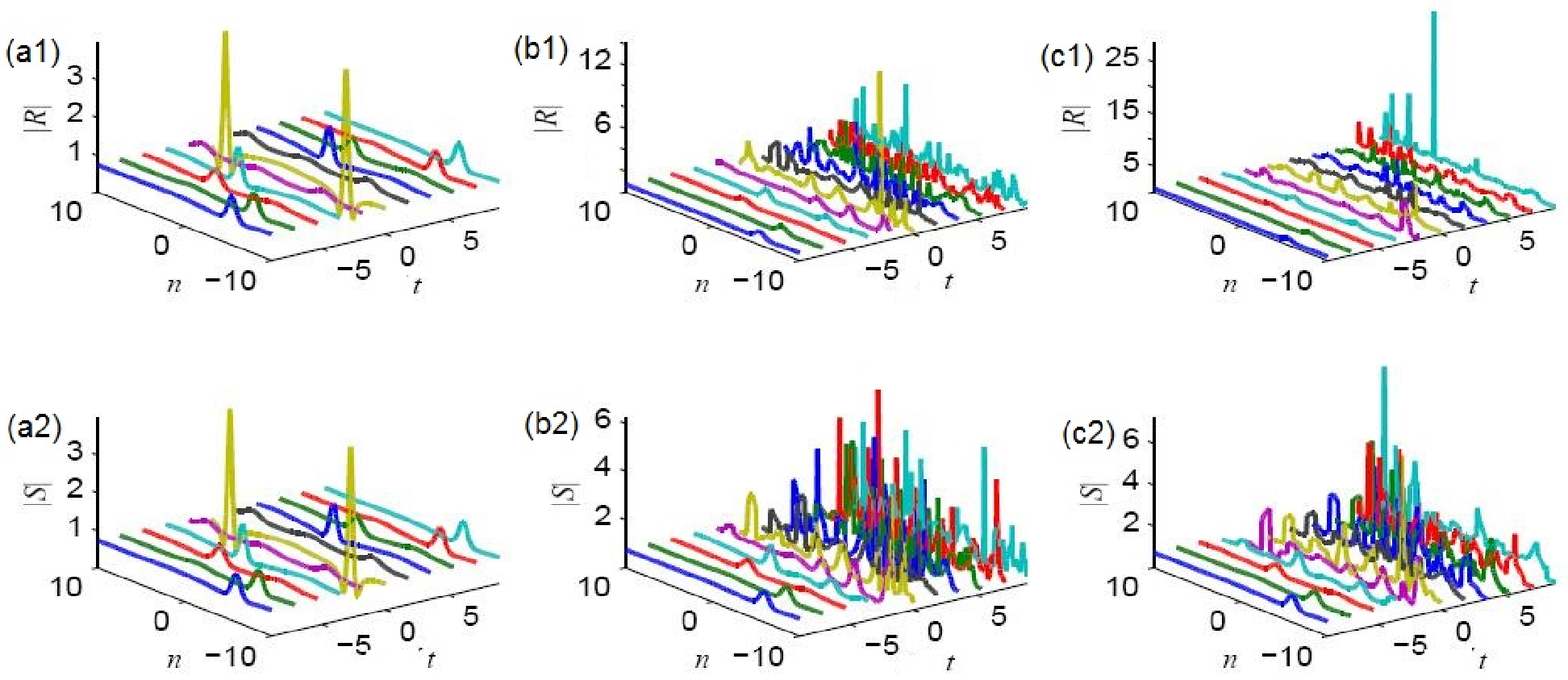}}}
\end{center}
\par
\vspace{-0.3in}
\caption{{\protect\small (color online). The third-order RW solutions (%
\protect\ref{rw3}) with $e_{1}=10,d_{1,2}=e_{2}=0$. The exact solution
(left); simulated evolution using exact solutions (\protect\ref{rw3}) as the
initial conditions (middle); the evolution initiated by the exact solution
perturbed by weak random noise with amplitude }${\protect\small 0.01}$%
{\protect\small \ (right).}}
\label{fig-rw31-noise}
\end{figure}

\begin{figure}[tbp]
\begin{center}
{\scalebox{0.56}[0.7]{\includegraphics{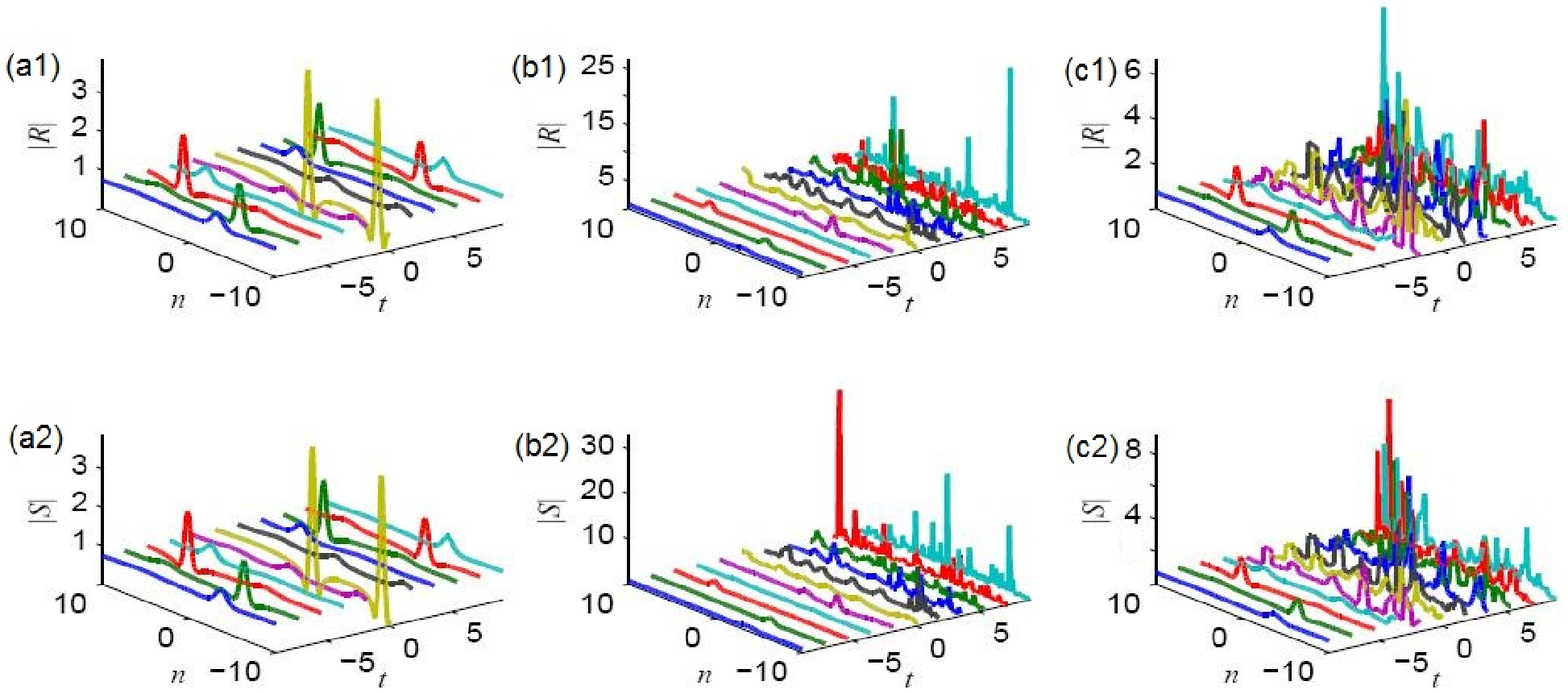}}}
\end{center}
\par
\vspace{-0.3in}
\caption{{\protect\small (color online). The third-order RW solutions (%
\protect\ref{rw3}) with $e_{2}=100,d_{1,2}=e_{1}=0$. The exact solution
(left); simulated evolution using exact solutions (\protect\ref{rw3}) as the
initial condition (middle); the evolution initiated by the exact solution\
perturbed by weak random noise with amplitude }${\protect\small 0.01}$%
{\protect\small \ (right).}}
\label{fig-rw32-noise}
\end{figure}

\textit{Dynamical behavior}---. We proceed to simulations of the evolution
of the third-order RW solutions $\widetilde{R}_{4}(n,t)$ and $\widetilde{S}%
_{4}(n,t)$ given by~Eq. (\ref{rw3}). We again consider regimes of the strong
and weak interaction, for which Figs.~\ref{fig-rw3-noise}-\ref{fig-rw32-noise}(a1) and (a2), respectively, show that the unperturbed
evolution agrees with the corresponding exact solutions~(\ref{rw3}) in a
short time interval. At $t>0$, the solution corresponding to the weak
interaction (Figs.~\ref{rw3-r}(b1,c1) and Figs.~\ref{rw3-s}(b1,c1)) exhibits
growing perturbations even in the absence of the initially added random
noise, which indicates instability (see Figs.~\ref{fig-rw3-noise}-\ref%
{fig-rw32-noise}(b1) and (b2)). This may be explained, as above, by the fact
that to the energy is spread in a larger domain, in comparison with ones in
Fig.~\ref{rw3-r}(a1) and Fig.~\ref{rw3-s}(a1). If weak random noise, with
amplitude $0.01$, is added to the initial conditions, see Figs.~\ref{rw3-r}
and~\ref{rw3-s}(a1,b1,c1), then the evolution of the third-order RW solution
in the weak-interaction regime exhibits strong instability, while the
perturbation growth is much slower in the strong-interaction regime, see
Fig.~\ref{rw3-r}(a1) and Fig.~\ref{rw3-s}(a1). Once again, these results may
be explained by the concentration of the energy around the origin in the $%
\left( n,t\right) $ plane under the action of the strong interaction, see
Figs.~\ref{fig-rw3-noise}-\ref{fig-rw32-noise}(a1), and by the spread of the
energy around several points in the case of the weak interaction, see Figs.~%
\ref{fig-rw3-noise}-\ref{fig-rw32-noise}(b1,c1).

%%%%%%%%%%%%%%%%%%%%%%%%%%%%%%%%%%%%%%%%%%%%%%%%%%%%
%%%%%%%%%%%%%%%%%%%%%%%%%%%%%%%%%%%%%%%%%%%%%%%%%%%%
%%%%%%%%%%%%%%%%%%%%%%%%%%%%%%%%%%%%%%%%%%%%%%%%%%%%
%%%%%%%%%%%%%%%%%%%%%%%%%%%%%%%%%%%%%%%%%%%%%%%%%%%%
%%%%%%%%%%%%%%%%%%%%%%%%%%%%%%%%%%%%%%%%%%%%%%%%%%%%
%%%%%%%%%%%%%%%%%%%%%%%%%%%%%%%%%%%%%%%%%%%%%%%%%%%%

Similarly, Theorem 2 with $N>4$ produces the higher-order RW solution of Eq.
(\ref{nls}), which can also generate the abundant discrete wave structures.

\section{Conclusion}

The subject of this work is the construction of families of higher-order
discrete RW (rogue-wave) states, which can be found in the integrable system
of coupled AL (Ablowitz-Ladik) equations. These equations may serve as
models for the dynamics of BEC trapped in a deep optical lattice. In the
work, we have presented a novel method for constructing generalized $(M,N-M)$%
-fold Darboux transforms (DTs) of the coupled AL system, which is based on
the fractional form of the respective determinants. Then, we apply the
generalized $(1,N-1)$-fold DTs to produce higher-order RW solutions. They
display a variety of patterns, including the rotating triangles and
pentagons, as well as quadrangle structures, which represent multi-RWs
in the system, and suggest shapes which such higher-order RW may assume in
other physically relevant models. The evolutions of the multi-RW solutions
was studied by means of systematic simulations, which demonstrate that the
tightly and loosely bound solutions are, respectively, nearly stable and
strongly unstable ones.

As an extension of this work, it may be interesting to look for similar
complex RW modes, by means of numerical methods, in more generic
nonintegrable discrete systems.

\vspace{0.1in}

\textbf{Acknowledgments} \vspace{0.01in}

The authors would like to thank the anonymous referees for their valuable comments and suggestions which help to
improve the manuscript. This work has been partially supported by the NSFC under Grants Nos.
11375030 and 11571346, and China Postdoctoral Science Foundation under Grant No. 2015M570161.
 \newline

\noindent \textbf{Appendix A}\newline
Explicit expressions for functions $\left\{ \phi ^{(2,3)}\right\} $ and $%
\left\{ \psi ^{(2,3)}\right\} $ in Eq. (\ref{e271}):\newline
$\phi ^{(2)}=-\frac{1}{78643200}(\frac{25}{16})^{n}e^{-\frac{55}{16}%
it}(55296000itd_{1}+49152000ie_{1}-20697600it^{2}+1966080in^{3}-2621440in-16934400int^{2}+589824in^{2}+455625it^{4}+58982400ine_{1}+589824in^{4}-3110400in^{2}t^{2}-819200i+3276800t+16711680nt+2211840tn^{3}-58982400nd_{1}-1944000nt^{3}+55296000te_{1}+11796480n^{2}t-7128000t^{3}-49152000d_{1}),
$\newline
\newline
$\psi ^{(2)}=-\frac{1}{13107200}(\frac{25}{16})^{n}e^{-\frac{5}{16}%
it}(-102400it+18432000ite_{1}-19660800ne_{1}+737280itn^{3}+3916800nt^{2}-648000int^{3}+2088960in^{2}t+3347200t^{2}+1474560itn-18432000td_{1}-19660800ind_{1}-151875t^{4}+131072n^{2}-196608n^{4}-1836000it^{3}+1036800n^{2}t^{2}),
$\newline
\newline
$\phi ^{(3)}=-\frac{1}{1006632960000}(\frac{25}{16})^{n}e^{-\frac{55}{16}%
it}(754974720000ine_{2}+20447232000tn^{3}+541900800000t^{2}d_{1}\newline
+212336640000in^{2}td_{1}+150994944in^{6}-199065600000int^{2}e_{1}-943718400000id_{1}^{2}+754974720in^{5}+754974720000intd_{1}+125829120in^{4}-17838080000it^{2}-4194304000in^{3}+188743680000in^{2}e_{1}+5326766080in+943718400000ie_{1}^{2}+3645000000t^{5}+629145600000ie_{2}+707788800000itd_{2}+2621440000i-3841982464in^{2}-93035520000nt^{3}-10485760000t+7549747200n^{2}t+150994944000ine_{1}-10485760000d_{1}-629145600000d_{2}+640942080000itd_{1}+75497472000in^{3}e_{1}-21676032000in^{3}t^{2}-541900800000it^{2}e_{1}-92318720000t^{3}-62208000000it^{3}d_{1}-78151680000in^{2}t^{2}-75497472000n^{3}d_{1}-754974720000nd_{2}-188743680000n^{2}d_{1}-27371520000n^{2}t^{3}+656100000nt^{5}+849346560tn^{5}+707788800000te_{2}+16135200000int^{4}+35283600000it^{4}-102515625it^{6}-62208000000t^{3}e_{1}-1887436800000e_{1}d_{1}-2488320000n^{3}t^{3}+7549747200n^{4}t-150994944000nd_{1}+640942080000te_{1}+754974720000nte_{1}+199065600000nt^{2}d_{1}+212336640000n^{2}te_{1}-95821824000int^{2}+1749600000in^{2}t^{4}-27561820160nt-1990656000in^{4}t^{2}+10485760000ie_{1}),
$\newline
\newline
$\psi ^{(3)}=-\frac{1}{503316480000}(\frac{25}{16})^{n}e^{-\frac{5}{16}%
it}(-88080384000ne_{1}+27869184000nt^{2}+37232640000n^{2}t^{2}-42091520000it^{3}-150994944n^{6}+3098250000it^{5}-1887436800000ie_{1}d_{1}+376012800000it^{2}d_{1}-88080384000ind_{1}+401080320000inte_{1}+199065600000int^{2}d_{1}+212336640000in^{2}te_{1}-401080320000ntd_{1}+15040512000n^{3}t^{2}+707788800000ite_{2}-1749600000n^{2}t^{4}-754974720000ne_{2}+199065600000nt^{2}e_{1}-13219200000nt^{4}-75497472000n^{3}e_{1}+4010803200in^{4}t-3853516800in^{2}t+4718592000itn^{3}-754974720000ind_{2}+1990656000n^{4}t^{2}-212336640000n^{2}td_{1}-2488320000in^{3}t^{3}+656100000int^{5}-55710720000int^{3}-62208000000it^{3}e_{1}+849346560itn^{5}+247726080000ite_{1}-21150720000in^{2}t^{3}-75497472000in^{3}d_{1}+983040000it-798720000t^{2}-23781600000t^{4}-247726080000td_{1}-707788800000td_{2}-943718400000e_{1}^{2}+943718400000d_{1}^{2}+102515625t^{6}+62208000000t^{3}d_{1}+376012800000t^{2}e_{1}+276824064n^{2}+503316480n^{4}-6590300160itn).
$


\begin{thebibliography}{99}
\bibitem{soliton} M. J. Ablowitz and P. A. Clarkson, \emph{Solitons,
Nonlinear Evolution Equations and Inverse Scattering} (Cambridge University
Press, Cambridge, 1990).

\bibitem{bec} L. Pitaevskii and S. Stringari, \emph{Bose-Einstein
Condensation} (Oxford University Press, Oxford, 2003).

\bibitem{Romania} V. S. Bagnato, D. J. Frantzeskakis, P. G. Kevrekidis, B. A. Malomed, and D. Mihalache, Rom. Rep. Phys. \textbf{67}, 5 (2015).

\bibitem{RW1} J. M. Dudley, F. Dias, M. Erkintalo, and G. Genty, Nature
Phys. \textbf{8}, 755 (2014).

\bibitem{RW2} N. Akhmediev, B. Kibler, F. Baronio, M. Beli\'{c}, W.-P.
Zhong, Y. Zhang, W. Chang, J. M. Soto-Crespo, P. Vouzas, P. Grelu, C.
Lecaplain, K. Hammani, S. Rica, A. Picozzi, M. Tlidi, K. Panajotov, A.
Mussot, A. Bendahmane, P. Szriftgiser, G. Genty, J. Dudley, A. Kudlinski, A.
Demircan, U. Morgner, S. Amiramashvili, C. Bree, G. Steinmeyer, C. Masoller,
N. G. R. Broderick, A. F. J. Runge, M. Erkintalo, S. Residori, U.
Bortolozzo, F. T. Arecchi, S. Wabnitz, C. G. Tiofack, S. Coulibaly, and M.
Taki, \emph{Viewpoint}, J. Optics \textbf{18}, 063001 (2016).

\bibitem{encyclopedia} B. A. Malomed, \emph{Nonlinear Schr\"{o}dinger
equations}, in: \emph{Encyclopedia of Nonlinear Science}, p. 639. Ed. A.
Scott. New York: Routledge, 2005.

\bibitem{review} B. A. Malomed, D. Mihalache, F. Wise, and L. Torner, J.
Optics B: Quant. Semicl. Opt. \textbf{7}, R53-R72 (2005); B. A. Malomed, D.
Mihalache, F. Wise, and L. Torner, \emph{Viewpoint}, J. Phys. B: At. Mol.
Opt. Phys. \textbf{49}, 170502 (2016).

\bibitem{bec2} R. Carretero-Gonz\'alez, D. J. Frantzeskakis, and P. G.
Kevrekidis, Nonlinearity \textbf{21}, R139 (2008).

\bibitem{bec3} Y. V. Kartashov, B. A. Malomed, and L. Torner, Rev. Mod.
Phys. \textbf{83}, 247 (2011).

\bibitem{Ma} W. X. Ma and M. Chen, Appl. Math. Comput. \textbf{215}, 2835 (2009).

\bibitem{f1} V. Ivancevic, Cognitive Computation \textbf{2}, 17 (2010); e-print, arXiv:0911.1834.

\bibitem{f2} Z. Yan, Commun. Theor. Phys. \textbf{54}, 947 (2010); e-print, arXiv:0911.4295.

\bibitem{f3} Z. Yan, Phys. Lett. A \textbf{375}, 4274 (2011).

\bibitem{nail} N. Akhmediev, J. M. Soto-Crespo, and A. Ankiewicz, Phys. Rev. A {\bf 80}, 043818 (2009).

\bibitem{nail2} A. Ankiewicz, Y. Wang, S. Wabnitz, and N. Akhmediev, Phys. Rev. E {\bf 89}, 012907 (2014).

\bibitem{yanpre10} Z. Yan, V. V. Konotop, and N. Akhmediev, Phys. Rev. E \textbf{82}, 036610 (2010).

\bibitem{yanpla10} Z. Yan, Phys. Lett. A \textbf{374}, 672 (2010).

\bibitem{guo1} B. L. Guo, L. M. Ling, Q. P. Liu, Phys. Rev. E \textbf{85}, 026607 (2012).

\bibitem{cm} S. Chen and D. Mihalache, J. Phys. A \textbf{48}, 215202 (2015).

\bibitem{yang12} Y. Ohta and J. Yang, Phys. Rev. E \textbf{86}, 036604 (2012).

\bibitem{yan-wen-pre} X. Wen, Y. Yang, and Z. Yan, Phys. Rev. E \textbf{92}, 012917 (2015).

\bibitem{yan-wen-chaos} X. Wen and Z. Yan, Chaos \textbf{25}, 123115 (2015).

\bibitem{yan-yang-chaos} Y. Yang, Z. Yan, and B. A. Malomed, Chaos \textbf{25}, 103112 (2015).

\bibitem{yan-non-15} Z. Yan, Nonlinear Dyn. {\bf 79}, 2515 (2015).

\bibitem{yan-chao-16} X. Y. Wen, Z. Yan, and Y. Yang,  Chaos {\bf 26}, 063123 (2016).

\bibitem{chen16} S. Chen, J. M. Soto-Crespo, F. Baronio, Ph. Grelu, and D. Mihalache, Opt. Express {\bf 24}, 15251 (2016).

\bibitem{yuan16}  F. Yuan, J. Rao, K. Porsezian, D. Mihalache, and J. S. He, Rom. J. Phys. {\bf 61}, 378 (2016).


\bibitem{Ablowitz} M. J. Ablowitz and J. F. Ladik, J. Math. Phys. \textbf{16}, 598 (1975).

\bibitem{nail1d} A. Ankiewicz, N. Akhmediev, and J. M. Soto-Crespo, Phys. Rev. E \textbf{82}, 026602 (2010).

\bibitem{nail2d} A. Ankiewicz, N. Akhmediev, and F. Lederer, Phys. Rev. E
\textbf{83}, 056602 (2011).

\bibitem{yang14} Y. Ohta and J. Yang, J. Phys. A: Math. Theor. \textbf{47},
255201 (2014).

\bibitem{wen-yan} X. Y. Wen and Z. Yan, to be submitted (2016).

\bibitem{optics} A. Szameit, F. Dreisow, M. Heinrich, S. Nolte, and A. A.
Sukhorukov, Phys. Rev. Lett. \textbf{106}, 193903 (2011).

\bibitem{Kuba} O. Dutta, M. Gajda, P. Hauke, M. Lewenstein, D.-S. Luhmann,
B. Malomed, T. Sowinski, and J. Zakrzewski, Rep. Prog. Phys. \textbf{78},
066001 (2015).

\bibitem{MY} B. A. Malomed and J. Yang, Phys. Lett. A \textbf{302}, 163
(2002).

\bibitem{wen1} X. Y. Wen, D. S. Wang and X. H. Meng, Rep. Math. Phys.
\textbf{72}, 349 (2013).

\bibitem{nail3} A. Ankiewicz, N. Devine, M. \"{U}nal, A. Chowdury, and N.
Akhmediev, J. Opt. \textbf{15}, 064008 (2013).

\bibitem{wen2} X. Y. Wen, J. Phys. Soc. Jpn. \textbf{81}, 114006 (2012).
\end{thebibliography}
\end{document}